\documentclass[aps,reprint,amsmath,amssymb,superscriptaddress,floatfix]{revtex4-2}

\usepackage{amsmath}
\usepackage{graphicx}

\usepackage{xcolor}
\usepackage{xspace}
\usepackage{dcolumn}
\usepackage{bm}

\begin{document}
\title{Quasiparticle spectroscopy in technologically-relevant niobium using London penetration
depth measurements}

\author{Sunil Ghimire}
\affiliation{Ames National Laboratory, Ames, IA 50011, U.S.A.}
\affiliation{Department of Physics \& Astronomy, Iowa State University, Ames, IA
50011, U.S.A.}

\author{Kamal R. Joshi}
\affiliation{Ames National Laboratory, Ames, IA 50011, U.S.A.}

\author{Amlan Datta}
\affiliation{Ames National Laboratory, Ames, IA 50011, U.S.A.}
\affiliation{Department of Physics \& Astronomy, Iowa State University, Ames, IA
50011, U.S.A.}

\author{Aidan Goerdt}
\affiliation{Ames National Laboratory, Ames, IA 50011, U.S.A.}

\author{Makariy A. Tanatar}
\affiliation{Ames National Laboratory, Ames, IA 50011, U.S.A.}
\affiliation{Department of Physics \& Astronomy, Iowa State University, Ames, IA
50011, U.S.A.}

\author{Deborah Schlagel}
\affiliation{Ames National Laboratory, Ames, IA 50011, U.S.A.}

\author{Matthew J. Kramer}
\affiliation{Ames National Laboratory, Ames, IA 50011, U.S.A.}

\author{Jayss Marshall}
\affiliation{Rigetti Computing, 775 Heinz Ave., Berkeley, CA 94710, U.S.A.}

\author{Cameron J. Kopas}
\affiliation{Rigetti Computing, 775 Heinz Ave., Berkeley, CA 94710, U.S.A.}

\author{Joshua Y. Mutus}
\affiliation{Rigetti Computing, 775 Heinz Ave., Berkeley, CA 94710, U.S.A.}

\author{Alexander Romanenko}
\affiliation{Fermi National Accelerator Laboratory, Batavia, Illinois 60510, U.S.A.}

\author{Anna Grassellino}
\affiliation{Fermi National Accelerator Laboratory, Batavia, Illinois 60510, U.S.A.}

\author{Ruslan Prozorov}
\email[Corresponding author: ]{prozorov@ameslab.gov}
\affiliation{Ames National Laboratory, Ames, IA 50011, U.S.A.}
\affiliation{Department of Physics \& Astronomy, Iowa State University, Ames, IA
50011, U.S.A.}

\date{\today}

\begin{abstract}
London penetration depth was measured in niobium foils, thin films, single crystals, and superconducting radio-frequency (SRF) cavity pieces cut out from different places. The low-temperature ($T<T_{c}/3$) variation, sensitive to the low-energy quasiparticles with states inside the superconducting gap, differs dramatically between different types of samples. With the help of phenomenological modeling, we correlate these different behaviors with known pair-breaking mechanisms and show that such measurements may help distinguish between different pair-breaking mechanisms, such as niobium hydrides and two-level systems (TLS). The conclusions also apply to SRF cavities when tracking the temperature-dependent quality factor and the resonant frequency. 
\end{abstract}
\maketitle

\section{Introduction}

Superconductors are characterized by unique properties that make them
particularly attractive for quantum computing \cite{Grezes2016,Kjaergaard2020,Huang2020,He2021,Siddiqi2021,Xiong2022,Ezratty2023}
and accelerator technologies \cite{padamsee2009,Anne_2022}. Niobium
is often used as at least some part of these technologies. This is
due in large part to its low resistivity at low temperatures, high
thermal conductivity, and highest among elements superconducting transition
temperature, $T_{c}\approx9.3$ K \cite{Stromberg1965,Finnemore1966,Daams1980,Bahte1998,Koethe2000,Prozorov2006a,Kozhevnikov2017,Liarte_2017}.
These attributes make niobium a popular choice for fabricating qubits
based on Josephson junctions \cite{wendin2017quantum}. Niobium is
a vital material in accelerator technology due to its capacity to
carry microwaves without significant losses. The high quality factor
(Q-factor), which signifies the efficiency of energy storage in a
resonator relative to energy loss, along with low surface resistivity
at relatively high magnetic fields, makes niobium an excellent choice
for superconducting radio frequency (SRF) cavities used in particle
accelerators \cite{padamsee2008,padamsee2009,Gurevich2012,Gurevich2017a,Ngampruetikorn2019,Ueki2022,Ueki2022a}
.

While a significant effort has been devoted to study properties of
niobium over years with first significant results appearing in the
late 1930s \cite{Daunt1937}, there are still fundamental aspects
that require further research. Measurement capabilities as well as
theoretical understanding of superconductors evolved immensely and
new studies bring novel results to this day. For example, based on
the first-principles microscopic theory of anisotropic superconducting
and normal state in Nb \cite{Zarea2022}, it was recently proposed
that, intrinsically, niobium is a type-I superconductor \cite{Prozorov2022}.
(Perhaps, all elemental superconductors are!) However,
in real samples and devices, disorder always tips the balance over
to the type-II side, but not too far from the boundary separating
these two regimes. In this situation, the electromagnetic response
is close to non-local since the coherence length and the London penetration
depth are comparable, of the order of 30-50 nm \cite{Prozorov2022}.

As all refractory metals, niobium has some physical-chemical issues
that complicate and sometimes impede its use in applications. One
of the most pressing issues in Nb SRF cavities is the so-called ``hydrogen
Q - disease'', a severe degradation of the quality factor, Q \cite{Barkov2012,Barkov2013}.
Niobium has significant affinity for hydrogen and can intake it even
from water and ambient moisture. At room temperature, small hydrogen
moves through niobium lattice as a free molecular gas. However, niobium
hydrides form upon cooling below 150-180 K and, depending on hydrogen
concentration, steric effects (volume mismatch) may irreversibly damage
initially perfect crystalline structure \cite{Barkov2012,Barkov2013}.
This damage is practically impossible to remove. Furthermore, in the
applications relying on the superconducting properties, the most important
part of any Nb structure is the surface layer where electromagnetic
field penetrates or supercurrent flows. When the Nb part is inevitably
exposed to air, a few nanometers thick layers of different niobium
oxides and sub-oxides form and, depending on their nature, may drastically
degrade device properties \cite{alex20}. Moreover, oxygen diffusing
deeper into the bulk may bind hydrogen forming two-level systems (TLS)
that create bound states at low energies, deep inside the superconducting
gap, which is extremely detrimental for quantum coherence. The TLS-related
losses represent a significant portion of the contemporary research
in applied superconductivity \cite{Burnett2016,Mueller2019,McRae20}.

Therefore, continuing studies of niobium's intrinsic properties,
in particular its electromagnetic response, are still needed to improve
qubits, accelerators, and other technologies that leverage the unique
traits of this metal. The ongoing significant effort in development
these technologies emphasizes the importance of such research, which
directly influences advancements in many sectors, including cryptography,
optimization, and high-energy physics \cite{preskill2018quantum}.

With so many factors that may affect the properties, hence the ultimate
performance of superconducting devices, the question is how to distinguish
between different mechanisms that create problems? For example, there
is no universal suppression of properties by generic hydrides. Their
influence depends on concentration, morphology, and size distribution.
Likewise, the influence of surface layers of oxides and hydrides depend
on their thickness, conductivity, magnetism and, of course, chemical
makeup \cite{An_oxide_2003,Burnett2016,Ngampruetikorn2019,Samsonova2021,Verjauw2021,Murthy22a}.

To (partially) address these questions, we performed precision measurements
of the London penetration depth using sensitive tunnel-diode resonator
technique. Seven representative samples were studied: different parts
of niobium SRF cavity; a thin film used in a superconducting transmon
qubit; commercial foil; and single crystals for comparison. The samples
were not treated in any special way and most contained the hydrides
from previous handling, but not all as we could observe directly in
low-temperature polarized microscope.

At the low temperatures, below roughly $T_{c}/3$, the superconducting
gap is constant and the temperature - dependent superfluid density
is determined by the quasiparticles that were created by different
non-thermal pair-breaking mechanisms, such as TLS or spin-flip scattering.
In order to understand the results, we modeled penetration depth using
Dynes model of superconductivity, first by itself with only two parameters,
pair-breaking and pair-conserving, $\Gamma$ and $\Gamma_{s}$, respectively,
and then extended to incorporate TLS into the total density of states.
As a result, we found a unique fingerprint of TLS in our measurements,
thus providing protocol on how to identify TLS and distinguish them
from other sources of pair-breaking scattering.

\begin{figure}[tbh]
\includegraphics[width=8.6cm]{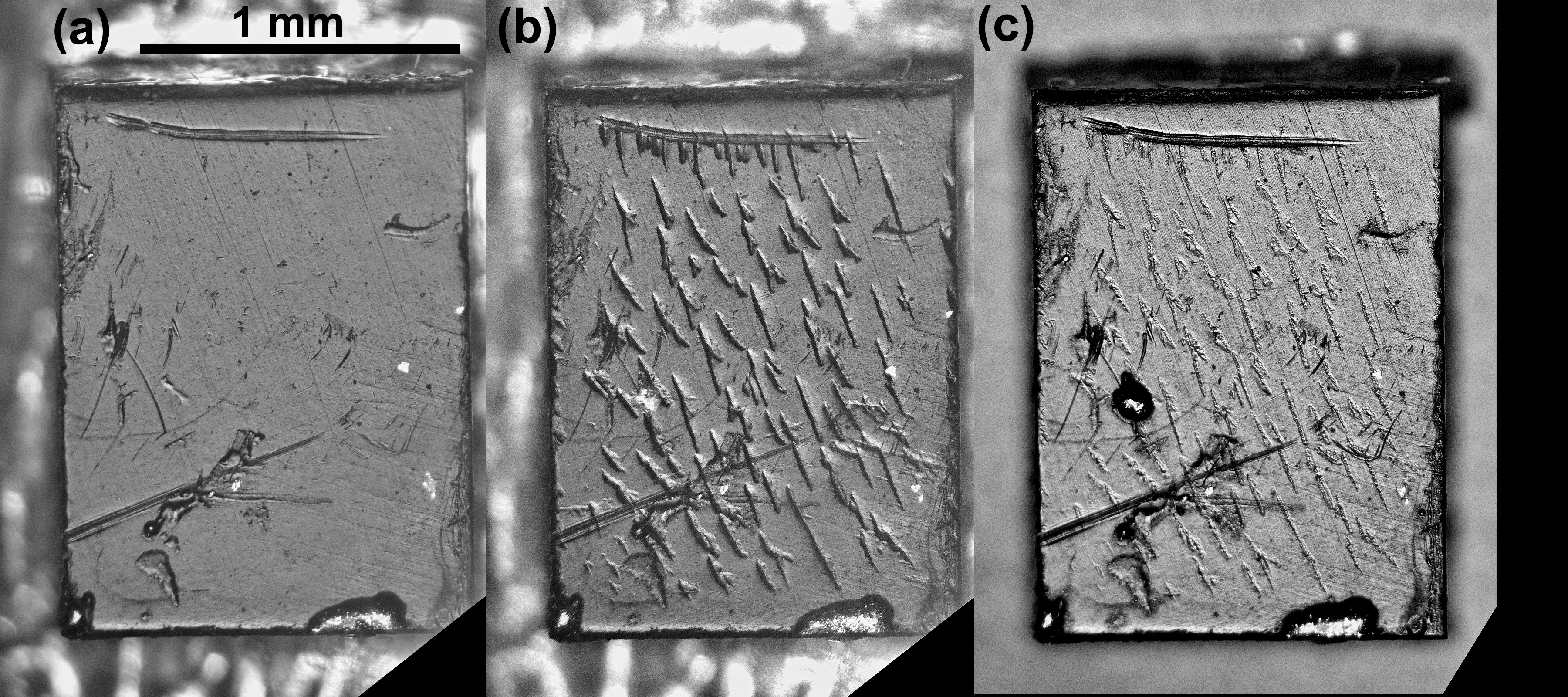} \caption{Polarized-light images of a {[}110{]} single crystal full of hydrogen.
(a) room-temperature image of the crystal that was never cooled down;
(b) the same crystal at 5 K showing profound hydrides outgrowths;
(c) room temperature image of the crystal warmed after cooling showing
scars - damage from the hydrides. These scars cannot be removed even
by heating the sample to sub-melting temperatures indicating significant
plastic deformation induced by the hydrides.}
\label{fig1:crystals}
\end{figure}

\section{Experimental}

\subsection{Samples}

Seven niobium samples (four different types) were used in this work.
Indeed, many samples of each type were measured during this study.
The sample types are: (1,2) Two single crystals, both cut from the
same large ingot and polished using water. They absorbed large amounts
of hydrogen. (3) Commercial Nb foil from Alfa Aesar, 250 $\mu$m thick,
99.98\% purity. It has been kept in a desiccator and showed no presence
of hydrides. (4) Sputtered 160 nm thick niobium film, also kept dry
and showing no obvious hydride formation. Identical films are used
to make transmon qubits \cite{Nersisyan2019a,Murthy22a}. (5,6,7)
Samples cut out from different places of a real SRF cavity. A surface
thermal map was constructed where surface temperature variations were
measured in different spots during the resonance. In some places,
called ``hot spots'' temperature rose by up to 1~K, whereas in
other, called ``cold spots'', changed only a little by 40~mK. Details
of this mapping and measurements are found elsewhere \cite{Alex09a}.
All samples were cut and dry-polished down to sub-mm size to fit in
our measurement setup.

\subsection{London penetration depth}

The London penetration depth was measured by using a sensitive tunnel
diode resonator (TDR) technique \cite{VanDegrift1975RSI,Prozorov2000,Prozorov2000a,Prozorov2006,Prozorov2011,Carrington2011,Prozorov2021}.
Essentially, TDR is a tank circuit where the sample is inserted into
a single-layer-of-turns inductor that produces a small, $H_{ac}<2~\text{\ensuremath{\mu}T}$,
AC magnetic field at around $f_{0}\approx14$ MHz. Connected in series
tunnel diode, biased to the regime of negative differential resistance,
compensates for losses in the circuit and for a certain impedance
matching conditions, the circuit starts resonating spontaneously,
usually below 70 K or so. If the diode and the circuit are well stabilized
and isolated, the resolution of the device is about 1 part per billion
resolving 0.01 Hz changes on top of 10 MHz main frequency. When a
sample responds magnetically, the total inductance changes leading
to a frequency shift, $\Delta f/f_{0}$, proportional to the samples
magnetic susceptibility, $\chi\left(T\right)$, with a sample-dependent
calibration constant, $\Delta f\left(T\right)/f_{0}=G\chi\left(T\right)$.
The susceptibility may then be converted to the London penetration
depth, using, $\left(1-N\right)\chi=\lambda/R\tanh\left(R/\lambda\right)-1$,
knowing demagnetizing factor, $N$ \cite{Prozorov2018}, and the effective
sample dimension, $R$ \cite{Prozorov2021}. Since at low temperatures
(practically below $0.8T_{c}$) the $\tanh$ term can be dropped and
temperature dependence of the susceptibility analyzed since $\chi\left(T\right)\sim\lambda\left(T\right)$.
Detailed description of the measurement procedure \cite{Prozorov2000a},
calibration \cite{Prozorov2000,Prozorov2018,Prozorov2021} and applications
\cite{Prozorov2006,Prozorov2011,Carrington2011} provide a complete
description of this unique technique capable of resolving sub-angstrom
changes in the London penetration depth, $\lambda\left(T\right)$,
in sub-mm sized crystals. Here we use it to study niobium samples
from different sources.

\begin{figure}[tbh]
\includegraphics[width=8cm]{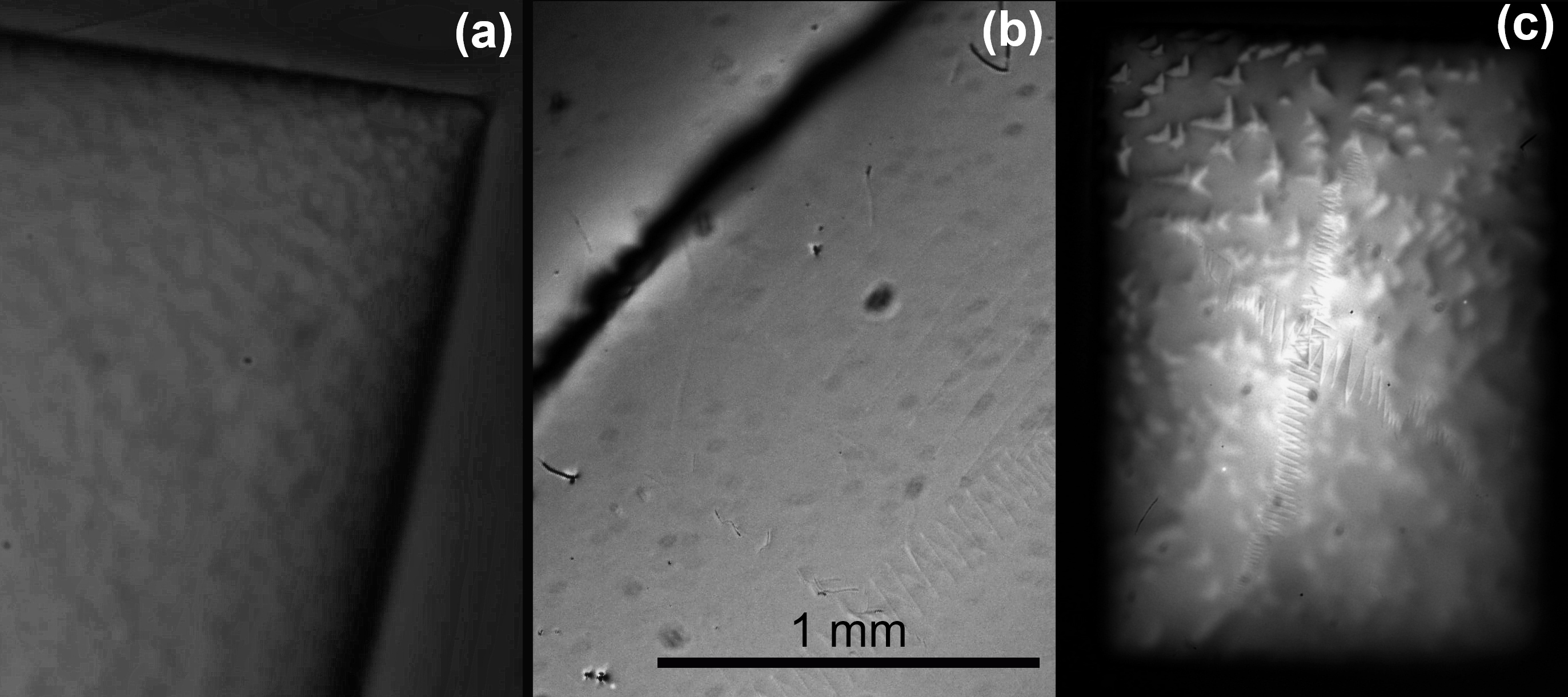} \caption{Magneto-optical Faraday images of three different cutouts from the SRF cavity. (a) cold spot from cavity's inner surface showing small
tubercles-like hydride structure; (b) a cutout 1 mm deep from cold
spot showing no hydrides; (c) hot spot from the surface with characteristic
boomerang-shaped large hydrides.}
\label{fig2:cavity}
\end{figure}

\subsection{Low-temperature optical and magneto-optical imaging}

The two-dimensional distribution of the magnetic induction was mapped
in real-time employing magneto-optical imaging using Faraday effect
in transparent ferrimagnetic indicators (bismuth-doped iron garnets)
placed on top of the samples. Details of the technique can be found
in our previous studies of Nb \cite{Young2005,Prozorov2006a}. The
closed-cycle flow-type optical $^{4}$He cryostat exposed the cooled
sample to an Olympus polarized-light microscope. The magnetic induction
on the sample surface polarizes in-plane magnetic moments in the indicator,
and the distribution of this polarization component along the light
propagation is visualized through double Faraday rotation. In the
images, only the magnetic field is visible due to a mirror sputtered
at the bottom of the indicator.

The same microscope with the cryostat were used for direct observation
in linearly polarized light. Due to polarization, all surface features
become of higher contrast, because they often cause some rotation
of the polarization plane upon reflection and since we work in (almost)
crossed polarizer/analyzer configuration.

\section{Results and Discussion}

\subsection{Optical measurements}

We start with optical characterization of the samples. Figure \ref{fig1:crystals}
shows polarized light imaging one of the {[}110{]} oriented (out of
the page direction) single crystals cut from a big piece in presence
of water, so it absorbed a significant amount of hydrogen. Panel (a)
shows this crystal at the room temperature, never cooled down. No
signature of surface features are seen. Panel (b) shows the same crystal
at 5 K revealing large hydride formations. In our studies, we found
that the hydrides tend to grow along principal directions and, therefore
elongated shape here is not surprising. The direction, however, would
depend on the strain, which is unknown. Figure~\ref{fig1:crystals}(c)
again shows a room-temperature image, but now after the cool-down.
Scars left by the hydrides are clearly visible. These significant
deformations and rapture of the previously perfect crystal would remain
even if the crystal is heated up only few degrees below melting point
of niobium. Therefore, the penetration depth data shown below were
collected on a crystal full of large hydrides. 

Next we examine three different cutouts from the SRF cavity. Two are
from the inner surface and one from the depth of 1 mm beneath. Figure
\ref{fig2:cavity} shows magneto-optical images of each part in the
remanent state of a trapped magnetic flux. (The sample is cooled in
a magnetic field to low temperature and the field is turned off.)
Importantly, sample surface is not visible, only trapped magnetic
flux distribution. Figure \ref{fig2:cavity} (a) shows the cold spot
(where temperature variation did not exceed 40 mK). There are hydrides
in shape of small tubercles. Panel (b) shows cutout from 1 mm depth
under the cold spot revealing clean surface without the hydrides indicating
that hydrogen distribution is highly non-uniform depth-wise. The third
panel (c), shows a hot spot (temperature rose more than 1 K in a resonating
cavity) revealing large hydrides similar to those observed in a single
crystal, Fig.\ref{fig1:crystals}.

We will now examine temperature-dependent superfluid density in these
samples. This measurement brings information about the superconducting
gap structure and possible pair-breaking mechanisms. In addition to
crystals and cavity cutouts, we also measured a thin film used in
the fabrication of transmons as well as commercial foil - to cover
all possible states of niobium samples. Thin films from the same batches
were characterized in great detail elsewhere \cite{Joshi2022arXiv}. 

\subsection{London penetration depth}

\begin{figure}[tbh]
\includegraphics[width=8cm]{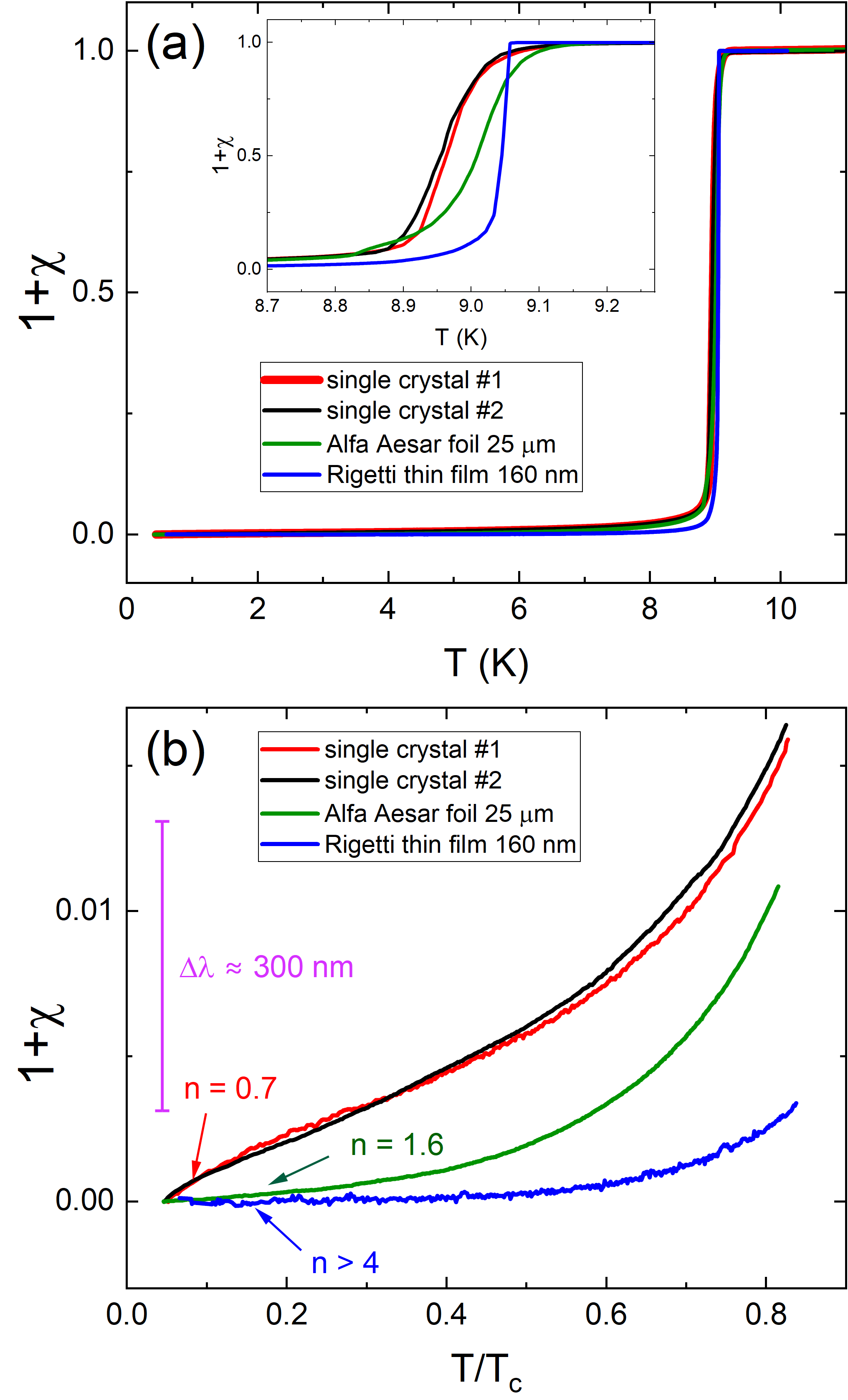} \caption{Magnetic susceptibility of two single crystals, Alfa Aesar foil and
160 nm Rigetti thin film. (a) main panel - full range, inset - zooming
into the superconducting transition. (b) Low-temperature variation
of London penetration depth fitted to the power-law, $\Delta\lambda\sim\left(T/T_{c}\right)^{n}$,
from the base temperature, $0.044T_{c}$ to $0.3T_{c}$ yielding the
values of the exponent $n$ indicated. Magenta bar show the corresponding
scale of $\lambda\left(T\right)$ variation in nm.}
\label{fig3:crystalsTDR} 
\end{figure}

The first set of measurements on two single crystals, Alfa Aesar foil
and Rigetti thin film is shown in Fig.\ref{fig3:crystalsTDR}. For
comparison between different samples the curves were normalized to
represent ideal magnetic susceptibility that starts at $\chi=0$ above
$T_{c}$ and reaches $\chi=-1$ at the low temperature. The detailed
shape of $\chi\left(T\right)$ is unaffected by this scaling. The
inset in panel (a) of Fig.\ref{fig3:crystalsTDR} zooms at the superconducting
transition showing the sharpest transition in Nb film and very similar
$T_{c}$ in two crystals. All these values are somewhat lower than
often quoted 9.3 K, likely due to disorder-induced pair-breaking in
this quite anisotropic material \cite{Zarea2022,p-irr-bridge}. Figure
\ref{fig3:crystalsTDR}(b) shows the low-temperature variation revealing
quite different behaviors of $\lambda\left(T\right)$. The power-law
fitting, $\Delta\lambda\sim\left(T/T_{c}\right)^{n}$ from the
base temperature, $0.044T_{c}$ to $0.3T_{c}$ produced the indicated
values of the exponent, $n$, which range from exponential attenuation,
$n>4$, to a convex downturn with $n=0.7$, which is extremely unusual
for any superconductor. In both single crystals we observe a downturn
below roughly 1.3 K, whereas this behavior is absent in a thin film
and a foil.

\begin{figure}[tbh]
\includegraphics[width=8cm]{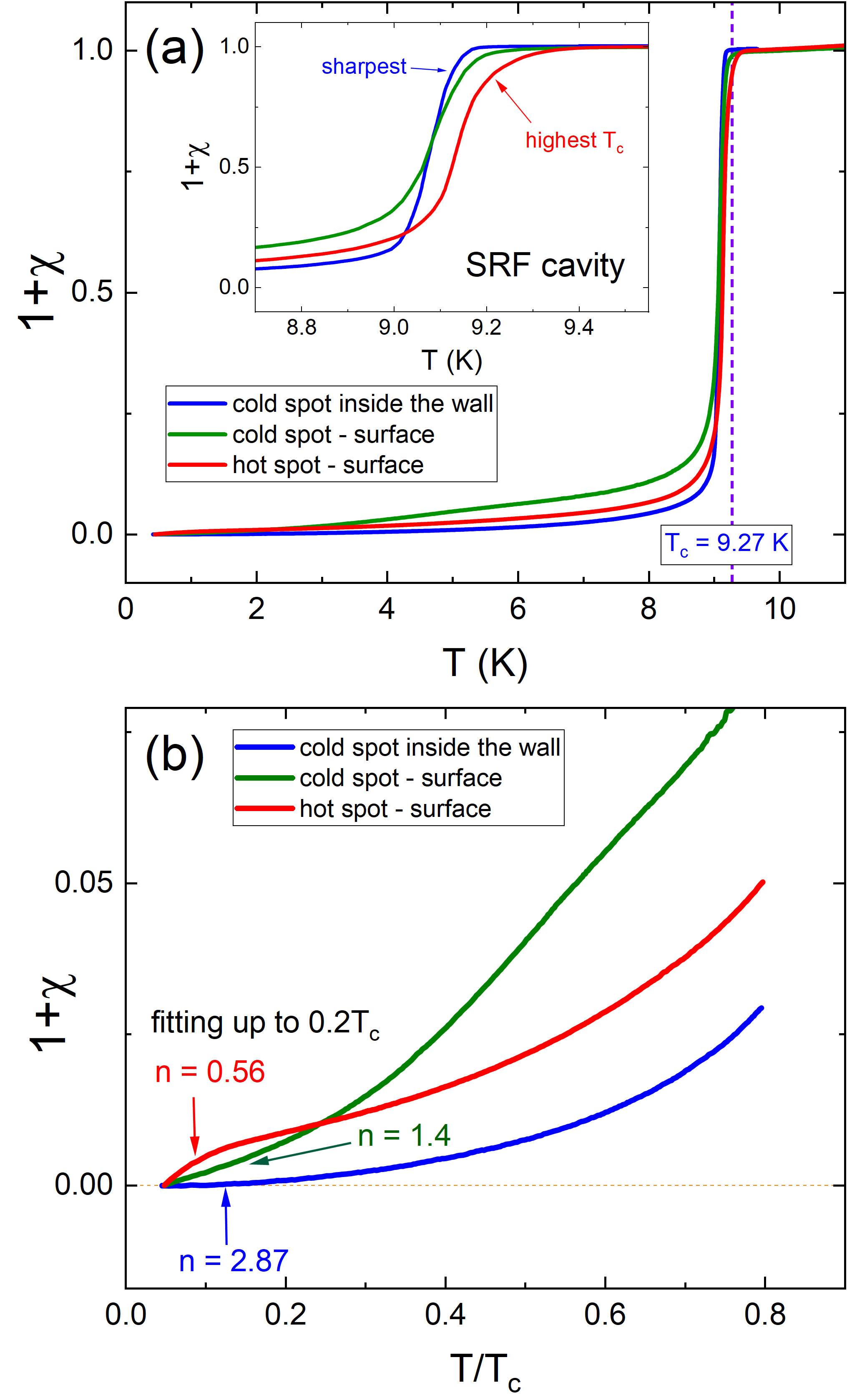} \caption{The information similar to Fig.\ref{fig3:crystalsTDR}, but for the
SRF cavity cutouts. Two from the surface, cold and hot spot, and the
third from the depth of 1 mm under cold spot. While the transitions
are higher, there is significant spread with widest transition width
of about 0.4 K in the hot spot sample and the narrowest in the inner
cold spot. (b) Low-temperature variation shows a distinct convex downturn,
which is very unusual for $\lambda\left(T\right)$ in any superconductor.
The inner cold spot shows saturation behavior consistent with a relatively
clean superconducting gap.}
\label{fig4:SRF} 
\end{figure}

The second set of data, shown in Fig.\ref{fig4:SRF}, presents the
results obtained in the SRF cavity cutouts in a way formally similar
to the first set, Fig.\ref{fig3:crystalsTDR}. Full temperature variation
is presented in the main panel of Fig.\ref{fig4:SRF}(a) and the inset
zooms into the transition region. While the transitions are higher
than in the first set, there is a more significant spread with the
widest transition width of about 0.4 K in the hot spot sample and
the narrowest in the inner cold spot. Panel (b) of Fig.\ref{fig4:SRF}
presents low-temperature variation, showing even more pronounced downward
curvature, but only for one sample that came from the cold spot. This
fact will be the key for the interpretation of these results. Such
distinct convex downturn is very unusual for $\lambda\left(T\right)$
in any (good and uniform) superconductor. The inner cold spot shows
saturation behavior consistent with a relatively clean superconducting
gap and the surface cold spot shows an intermediate behavior.

In principle, such behavior of the single crystals in Fig.\ref{fig3:crystalsTDR}
and the hot spot sample in Fig.\ref{fig4:SRF} could be due to the
proximity effect when superconductivity is induced in the surface
metallic layer and the overall diamagnetic screening increases thus
leading to the downturn in susceptibility. This was shown directly
using TDR technique in MgB$_{2}$ wires that has excess magnesium
on the surface \cite{Prozorov_Proximity}. Pambianchi \emph{et al.} measured
effective penetration depth in proximity-coupled Nb/Al bilayer films
where they observed power law with exponent n$\leq$ 1 distinctly
different from the exponential behavior of Nb \cite{Analage_1995}.
However, in both works, the effect required a quite thick layer of
another normal metal on niobium surface. It was suggested that niobium
hydride phase, Nb$_{4}$H$_{3}$, which is also a candidate own superconductivity
with a critical temperature of 1.2 K \cite{roth1990suppression}.
However, our niobium samples are exposed to air and poorly or non-conducting
niobium oxides are formed on the surface thus eliminating the possibility
of the formation of a good metallic layer. The oxides are very robust
and withstand heating almost to melting temperature of Nb \cite{An_oxide_2003,berti2023scanning}.
Careful atom probe and transmission electron microscopy (TEM) studies showed no evidence of the metallic hydride phase on sample surfaces. In case of cavity cutouts,
why would these layers only form over hot spots? Finally, the most
important point is that proximity effect leads to the enhanced diamagnetism
and the downturn is the departure from the original $\lambda(T)$
curve downward upon cooling. What we observe is opposite - the whole curve
is shifted up indicating significant contamination of the superconducting
gap with quasiparticle states rather than additional screening. Still, keeping in mind this
possibility, we now turn to a more general analysis of the obtained
results. 

\section{Theoretical analysis}

\subsection{General remarks}

Here we present a simple phenomenological analysis where we use a
semiclassical connection between London penetration depth and the
density of states. The results were verified by Matsubara summation
formalism. Being phenomenological does not mean being qualitative.
We use self-consistency equation to calculate the order parameter
and then the superfluid density, and obtain quantitative prediction
for temperature-dependent penetration depth. We follow the semiclassical
approach introduced by Chandrasekhar et al. \cite{Chandrasekhar1993}.
Some examples of using this theory for different superconductors can
be found in Ref.\cite{Prozorov2006}. The model gives direct connection
between the superfluid density, $n_{s}\left(T\right)$, and the density
of states, $N\left(E\right)/N_{n}$ normalized by its value at the
Fermi level in the normal state, $N_{n}$. Here the energy of Bogolubov
quasiparticles is, $E=\sqrt{\epsilon^{2}+\Delta^{2}}$ and the
normal metal band energy, $\epsilon$, is measured from the Fermi
level. The response of supercurrent to a vector potential, $\mathbf{J=\mathbb{-R}A}$,
is determined by the so-called response tensor, $\mathbb{R}$, which
consists of two parts, diamagnetic and paramagnetic. The full expression
involves the average over the Fermi surface of a possibly anisotropic
gap function. In case of niobium we can safely use spherical Fermi
surface and constant gap (although the gap is somewhat anisotropic
\cite{Zarea2022}. Then the superfluid density, normalized on its
value in the clean case (which is the total electron density of normal
metal), reads:

\begin{equation}
n_{s}=1+2\intop_{0}^{\infty}\frac{\partial f\left(E\right)}{\partial E}\frac{N\left(E\right)}{N_{n}}dE\label{eq:ns}
\end{equation}
where the order parameter enters the density of states as,

\begin{equation}
\frac{N\left(E\right)}{N_{n}}=\frac{E}{\sqrt{E^{2}-\Delta^{2}}}\label{eq:DOSclean}
\end{equation}

The derivative of the Fermi function is, 
\begin{equation}
\frac{\partial f\left(E\right)}{\partial E}=-\frac{1}{4t}\text{sech}^{2}\left(\frac{E}{2t}\right)\label{eq:dfdE}
\end{equation}
\noindent where $t=T/T_{c}$ (note that this is actual superconducting transition
temperature that may be lower than the ``clean'' value that we denote
$T_{c0}$) and quasiparticle energy is dimensionless in units of $k_{B}T_{c}$
and we set $k_{B}=1$ everywhere. The quantity of interest, the London
penetration depth is then obtained from Eq.\ref{eq:ns} as,

\begin{equation}
\frac{\lambda\left(T\right)}{\lambda_{0}}=\frac{1}{\sqrt{n_{s}}}\label{eq:lambda}
\end{equation}
where $\lambda_{0}$ is a clean-limit London penetration depth. Therefore,
with scattering, this ratio will be greater than 1 even at $T=0$.

Importantly, this approach does not specify how the order parameter
is obtained. In the original papers of the semiclassical approach
where very clean superconductors were considered \cite{Einzel1986},
it was sufficient to use the so-called Einzel ansatz to analytically
represent the gap with high accuracy \cite{Einzel2003}. This is insufficient
for our purposes as we would like to explore ``contaminated'' gaps,
thus we will use the self-consistent solutions for $\Delta\left(T,\Gamma\right)$,
where $\Gamma$ is pair-breaking scattering rate \cite{Gurevich2017a}.

\subsection{Dynes superconductors}

The simplest way to introduce pair-breaking into the density of states
is to consider Dynes superconductors. There are several comprehensive
works with a detailed analysis of the Dynes model and its applicability
to realistic and quite complicated cases. A full set of all thermodynamic
and transport parameters was calculated and is available for reference.
We follow to extensive coverage of the Dynes model by Gurevich and
Kubo \cite{Gurevich2023,Gurevich2017,Kubo2020,Kubo2020a} and Herman
and Hlubina \cite{Herman2016,Herman2017,Herman2018}. In this model,
instead of Eq.\ref{eq:DOSclean}, the density of states is given by,

\begin{equation}
\frac{N\left(E\right)}{N_{n}}=\Re\left[\frac{E+i\Gamma}{\sqrt{\left(E+i\Gamma\right)^{2}-\Delta^{2}}}\right]\label{eq:DOSDynes}
\end{equation}

Dynes superconductors are gapless. At $E=0$, Eq.\ref{eq:DOSDynes}
gives $N\left(E\right)=N_{n}\Gamma/\sqrt{\Gamma^{2}+\Delta^{2}}$.
However, for small rates (usually found in tunneling experiments \cite{Lechner_PRA_2020,berti2023scanning}),
$\Gamma\ll\Delta$, one obtains quite useful estimates. In general
the meaning of this phenomenological parameter is that $\hbar/\Gamma$
is the quasiparticle lifetime. It also has meaning of pair-breaking
scattering time, because Dynes $\Gamma$ is analogous to Abrikosov-Gor'kov
(AG) scattering parameter \cite{AbrikosovGorkov1960ZETF}, 
\begin{equation}
\rho=\frac{\hbar}{2\pi T_{c}\tau}=\frac{\Gamma}{2\pi}
\label{scatrate}
\end{equation}
(actually, in the original paper, AG used $\rho$ without ``2''
in the denominator), and the $T_{c}$ for Dynes superconductor is
obtained from a classical AG formula \cite{Gurevich2017},

\begin{equation}
\psi\left(\frac{\Gamma}{2\pi t_{c}}+\frac{1}{2}\right)-\psi\left(\frac{1}{2}\right)+\ln\left(t_{c}\right)=0
\label{tcAG}
\end{equation}
\noindent where $t_{c}=T_{c}/T_{c0}$ and $\psi\left(x\right)$ is digamma function.
The transition temperature becomes zero when $\Gamma_{crit}=0.8819$.

\begin{figure}[tbh]
\includegraphics[width=8cm]{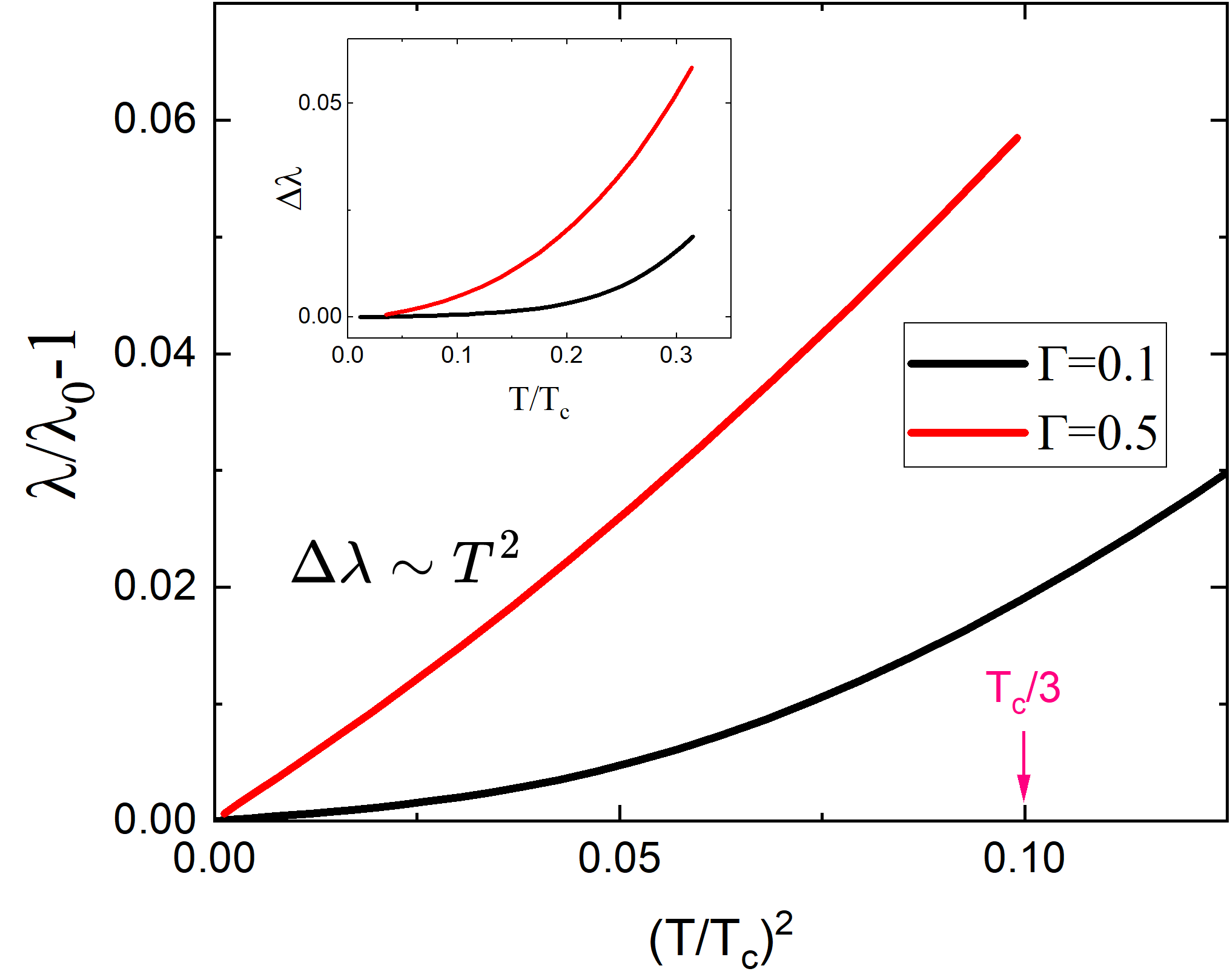} \caption{The change of the London penetration depth, calculated using Eqs.
\ref{eq:self-cons},\ref{eq:DOSDynes},\ref{eq:ns} and \ref{eq:lambda}.
(Here we put $\lambda_{0}=1$). At larger scattering rates it approaches
quadratic temperature dependence as predicted by Gurevich and Kubo
\cite{Gurevich2003}.}
\label{fig5:DynesLambda} 
\end{figure}

To calculate the superfluid density, we need to solve the
self-consistency equation for the order parameter written in terms
of Matsubara sum \citep{Gurevich2017},

\begin{equation}
\log(t)-\sum_{n=0}^{n_{max}}\left(\frac{2\pi t}{\sqrt{\Delta^{2}+\left(\Gamma+2\pi t\left(n+\frac{1}{2}\right)\right)^{2}}}-\frac{2}{2n+1}\right)=0\label{eq:self-cons}
\end{equation}

For each value of $\Gamma$ and for each temperature, $t$, Eq.\ref{eq:self-cons}
yields $\Delta\left(t,\Gamma\right)$. The number of terms, $n_{max}$
depends on the required accuracy and the lower the temperature, the
more terms are needed. Here we used $n_{max}=10000$.

Now we are set to calculate the London penetration depth as function
of temperature for different values of $\Gamma$ and compare with
the experiment. One difficulty is that the formulation in form of
Eq.\ref{eq:ns} does not include the effect of pair-conserving scattering,
which nevertheless affects the superfluid density. For Dynes superconductor
this is solved by calculating superfluid density using Matsubara sum
as introduced by Herman and Hlubina \citep{Herman2017},

\begin{equation}
n_{s}=2\pi t\sum_{n=0}^{n_{max}}\frac{\Delta^{2}}{\Omega_{n}^{2}(\Omega_{n}+\Gamma_{s})}\label{eq:nsMatsubara}
\end{equation}
where $\Omega_{n}=\sqrt{\Delta^{2}+(\Gamma+\pi t(2n+1))^{2}}$.
Here two scattering parameters are present, the pair-breaking $\Gamma$
that also enters the original Dynes equation, Eq.\ref{eq:DOSDynes}
and pair-conserving scattering rate, $\Gamma_{s}$.

Figure \ref{fig5:DynesLambda} shows the variation of the London penetration
depth, calculated using Eqs.~\ref{eq:self-cons},\ref{eq:DOSDynes},\ref{eq:ns}
and \ref{eq:lambda}. (Here $\lambda_{0}=1$). Identical results
are obtained from Eq.\ref{eq:nsMatsubara} with $\Gamma_{s}=0$. At
fairly small rates, the behavior is sub-quadratic, but clearly non-exponential
indicating significant deviation from the clean (exponential) case.
At larger scattering rates penetration depth approaches quadratic
temperature dependence as predicted by Gurevich and Kubo \cite{Gurevich2003}.
Compared with our experimental results, such behavior would be consistent
with the Alfa Aesar foil, Fig.\ref{fig3:crystalsTDR} and, perhaps
to some extend with the cold spot shown in Fig.\ref{fig4:SRF}.

\begin{figure}[tb]
\includegraphics[width=8.5cm]{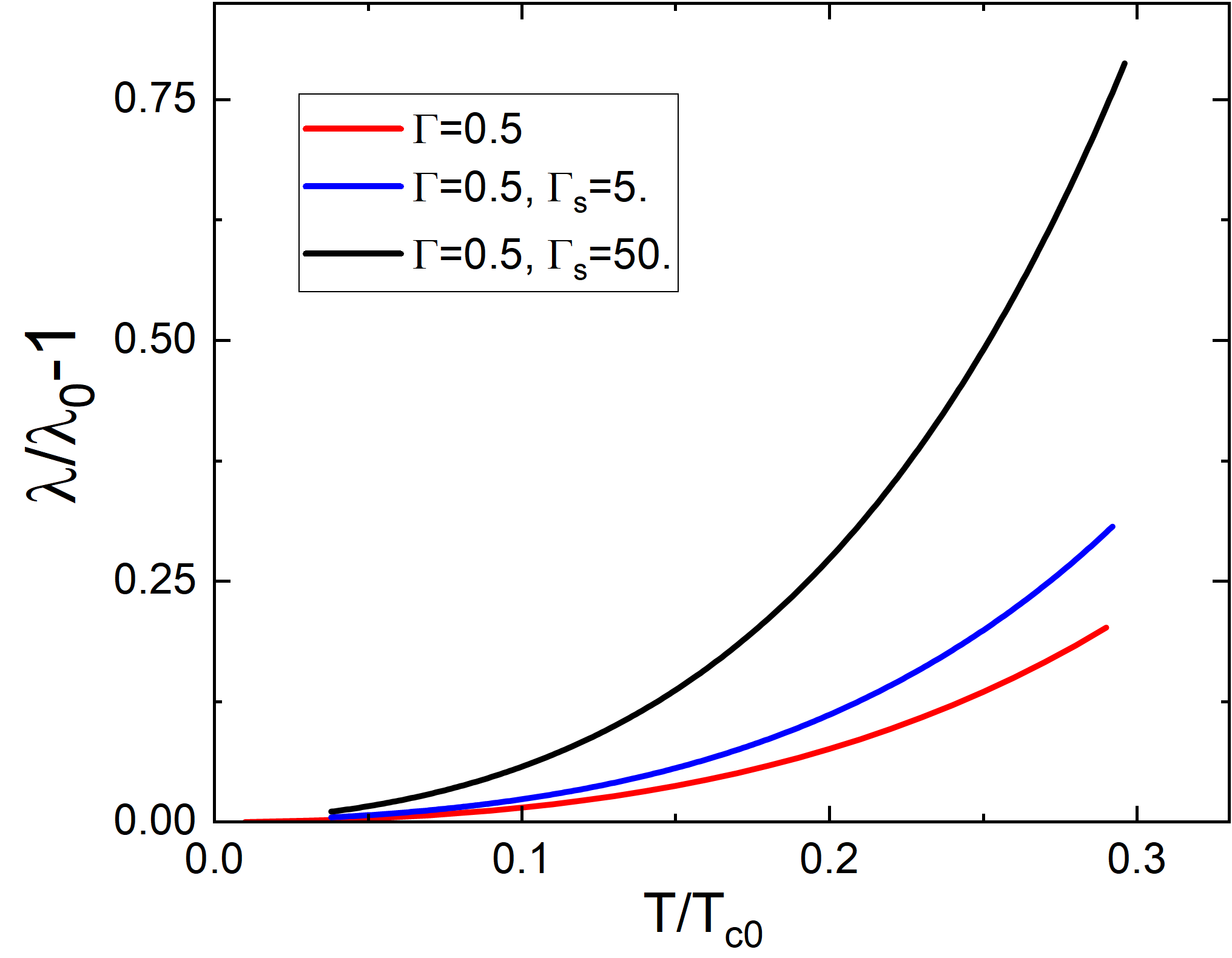} \caption{Temperature dependent London penetration depth, $\Delta\lambda(T)$,
calculated for a fixed Dyson pair-breaking $\Gamma=0.5$ and $\Gamma_{s}=0,5,50$.
While, expectedly, the slope becomes steeper, the behavior remains
close to quadratic (note that, unlike Fig.\ref{fig5:DynesLambda},
this is linear scale.)}
\label{fig6:DynesGammas} 
\end{figure}

Still, we are not even close to reproducing the downturn. Perhaps,
pair-conserving $\Gamma_{s}$ will help? Figure \ref{fig6:DynesGammas}
shows penetration depth for a fixed Dyson pair-breaking, $\Gamma=0.5$,
and pair-conserving, $\Gamma_{s}=0,5,50$ (unlike $\Gamma$, the pair-conserving
$\Gamma_{s}$ can assume any values, but dirty behavior starts with
the values above 1, so 50 is an exaggeration to see the difference.
While the temperature variation became much steeper, the functional
form remains sub-quadratic. To explore other possibilities we turn
to the two-level systems.

\subsection{Two-level systems and penetration depth}

\begin{figure}[tb]
\includegraphics[width=8.5cm]{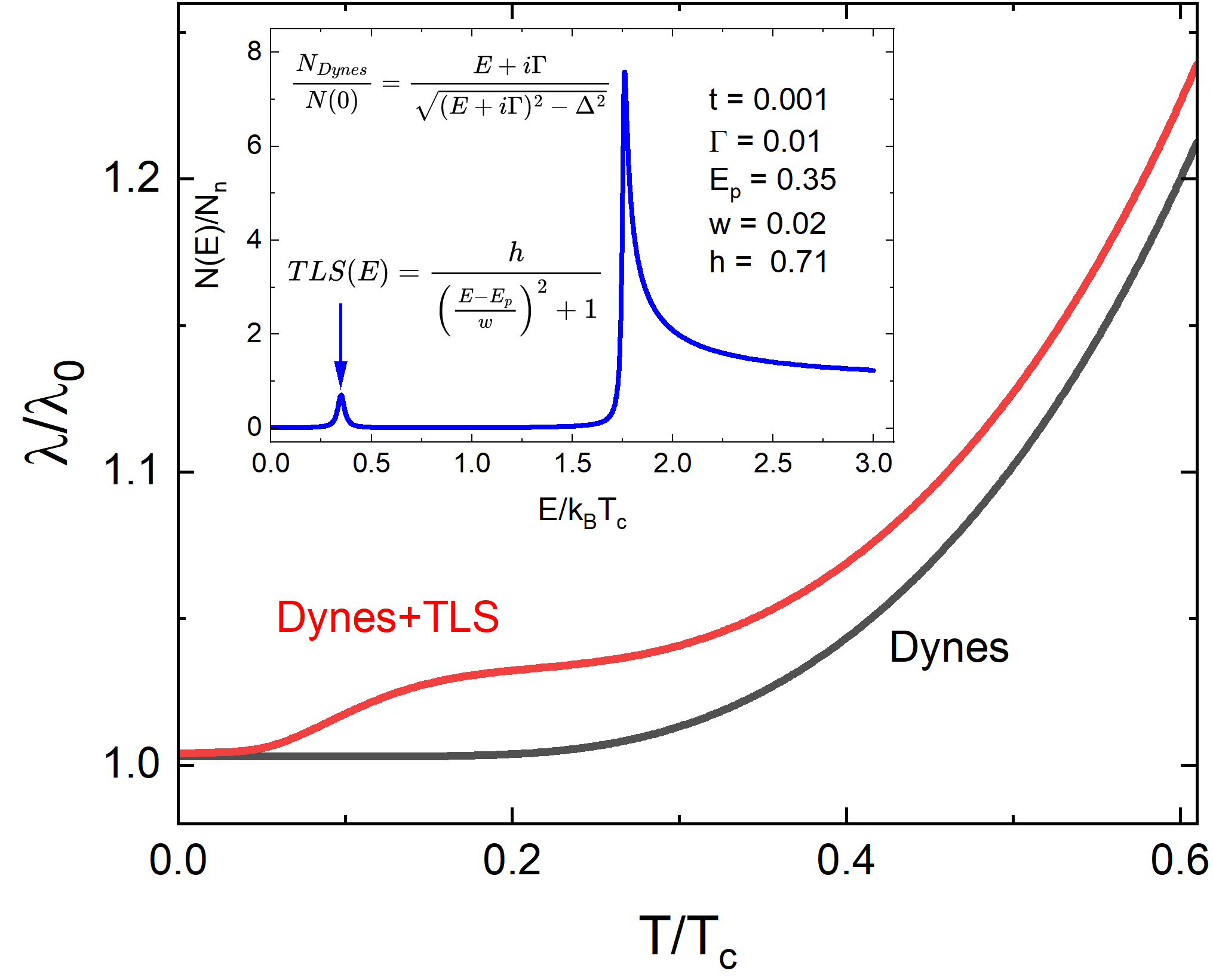} \caption{Temperature dependent London penetration depth, calculated using Eq.\ref{eq:ns}
with (red line) and without (black line) TLS contribution to the density
of states. The total $N\left(E\right)/N_{n}$ is shown in the inset.
A relatively small peak, with the parameters indicated, causes such
a dramatic change in the penetration depth. Importantly, it naturally
produces the downturn in the penetration depth. }
\label{fig7:TLSlambda} 
\end{figure}

\begin{figure}[tb]
\includegraphics[width=8.5cm]{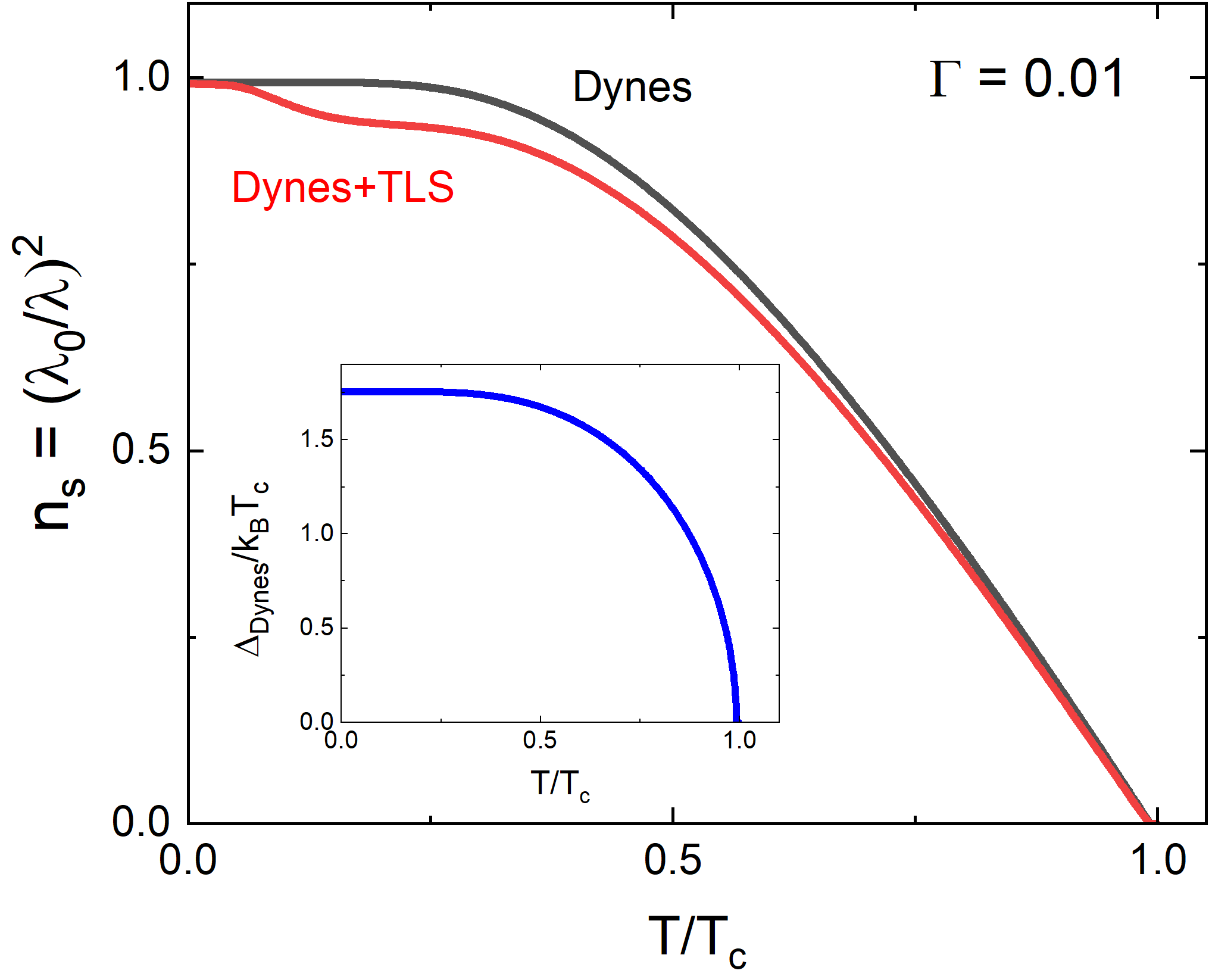} \caption{Temperature dependent superfluid density with and without TLS. The Dynes curve was computed for small $\Gamma=0.01$ and the TLS parameters
are shown in the inset in Fig.\ref{fig7:TLSlambda}. A pronounced
depression at low temperatures may explain various issues appearing
upon deep cooling. Inset shows self-consistent order parameter as
function of temperature.}
\label{fig8:TSLns} 
\end{figure}

There is a significant interest in TLS, which are intensely studied
theoretically and experimentally. Here we are only interested in their
possible effect on the London penetration depth. For that, it is sufficient
to model TLS as small peaks deep inside the superconducting energy gap in the density of states. Since the superfluid density is the integral over all energies, we do not think the precise details of the TLS peak may drastically change the outcome. Here we
model the TLS density of states as a Lorentzian,

\begin{equation}
TLS(E)=\frac{h}{\left(\frac{E-E_{p}}{w}\right)^{2}+1}\label{eq:TLSDOS}
\end{equation}
so that this peak of height $h$ and width $w$ is located at energy
$E_{p}$ inside the Dynes gap. An example is shown in the inset in
Fig.\ref{fig7:TLSlambda}. The total DOS is obtained by adding Eq.\ref{eq:DOSDynes}
and Eq.\ref{eq:TLSDOS}. Then Eq.\ref{eq:ns} is used to calculate
the superfluid density shown in Fig.\ref{fig8:TSLns}. The black curves
on both figures are the results for Dynes superconductor with small
$\Gamma=0.01$ showing close to isotropic s-wave BCS classical exponential
attenuation at low temperatures. The introduction of a small TLS peak
changes situation drastically. This is because its location deep inside
the gap. According to Eq.\ref{eq:ns}, the reduction of the superfluid
density is proportional to the product of DOS and derivative of the
Fermi function. The latter is steep at low temperatures and when it
reaches even a small peak, the result is significant.

Clearly, this mechanism naturally explains our observations of the
downturn in the penetration depth. While the hydrides are clearly
not good for applications because they increase overall surface impedance
and lead to significant dissipation, the downturn is not their signature.
Further research is needed to clarify the nature of TLS which is much
more complicated than used here and involved dynamic processes, we
believe our measurements and analysis presented a complimentary way
to identify TLS in niobium based applications.

\section{Conclusions}

Dyson pair-breaking $\Gamma$ changes temperature variation from exponential
to the power law, but not faster than quadratic behavior. The pair-conserving
$\Gamma_{s}$ only increases the rate of change, but does not significantly
alter the functional form of $\lambda(T)$. The two-level systems
residing inside the gap have profound effect changing exponential
attenuation to sub-linear downward concave curvature of $\lambda(T)$.
We emphasize that this was a limited study designed to probe a variety
of different types of samples. Systematic measurements of each type
are needed to identify the actual cause behind the TLS-like behavior. 

\begin{acknowledgments}
We thank James Sauls, John Zasadzinski, Alex Gurevich and Maria Iavarone
for useful discussions. This work was supported by the U.S. Department
of Energy, Office of Science, National Quantum Information Science
Research Centers, Superconducting Quantum Materials and Systems Center
(SQMS) under contract number DE-AC02-07CH11359. The research was performed
at the Ames National Laboratory, operated for the U.S. DOE by Iowa State University
under contract \# DE-AC02-07CH11358. 
\end{acknowledgments}


\begin{thebibliography}{68}%
\makeatletter
\providecommand \@ifxundefined [1]{%
 \@ifx{#1\undefined}
}%
\providecommand \@ifnum [1]{%
 \ifnum #1\expandafter \@firstoftwo
 \else \expandafter \@secondoftwo
 \fi
}%
\providecommand \@ifx [1]{%
 \ifx #1\expandafter \@firstoftwo
 \else \expandafter \@secondoftwo
 \fi
}%
\providecommand \natexlab [1]{#1}%
\providecommand \enquote  [1]{``#1''}%
\providecommand \bibnamefont  [1]{#1}%
\providecommand \bibfnamefont [1]{#1}%
\providecommand \citenamefont [1]{#1}%
\providecommand \href@noop [0]{\@secondoftwo}%
\providecommand \href [0]{\begingroup \@sanitize@url \@href}%
\providecommand \@href[1]{\@@startlink{#1}\@@href}%
\providecommand \@@href[1]{\endgroup#1\@@endlink}%
\providecommand \@sanitize@url [0]{\catcode `\\12\catcode `\$12\catcode
  `\&12\catcode `\#12\catcode `\^12\catcode `\_12\catcode `\%12\relax}%
\providecommand \@@startlink[1]{}%
\providecommand \@@endlink[0]{}%
\providecommand \url  [0]{\begingroup\@sanitize@url \@url }%
\providecommand \@url [1]{\endgroup\@href {#1}{\urlprefix }}%
\providecommand \urlprefix  [0]{URL }%
\providecommand \Eprint [0]{\href }%
\providecommand \doibase [0]{https://doi.org/}%
\providecommand \selectlanguage [0]{\@gobble}%
\providecommand \bibinfo  [0]{\@secondoftwo}%
\providecommand \bibfield  [0]{\@secondoftwo}%
\providecommand \translation [1]{[#1]}%
\providecommand \BibitemOpen [0]{}%
\providecommand \bibitemStop [0]{}%
\providecommand \bibitemNoStop [0]{.\EOS\space}%
\providecommand \EOS [0]{\spacefactor3000\relax}%
\providecommand \BibitemShut  [1]{\csname bibitem#1\endcsname}%
\let\auto@bib@innerbib\@empty
\bibitem [{\citenamefont {Gr{\`{e}}zes}(2016)}]{Grezes2016}%
  \BibitemOpen
  \bibfield  {author} {\bibinfo {author} {\bibfnamefont {C.}~\bibnamefont
  {Gr{\`{e}}zes}},\ }\href {https://doi.org/10.1007/978-3-319-21572-3} {\emph
  {\bibinfo {title} {Towards a Spin-Ensemble Quantum Mem. Supercond. Qubits
  Des. Implement. Write, Read Reset Steps}}},\ Springer Theses\ (\bibinfo
  {publisher} {Springer International Publishing},\ \bibinfo {year} {2016})\
  pp.\ \bibinfo {pages} {1--231}\BibitemShut {NoStop}%
\bibitem [{\citenamefont {Kjaergaard}\ \emph {et~al.}(2020)\citenamefont
  {Kjaergaard}, \citenamefont {Schwartz}, \citenamefont {Braumüller},
  \citenamefont {Krantz}, \citenamefont {Wang}, \citenamefont {Gustavsson},\
  and\ \citenamefont {Oliver}}]{Kjaergaard2020}%
  \BibitemOpen
  \bibfield  {author} {\bibinfo {author} {\bibfnamefont {M.}~\bibnamefont
  {Kjaergaard}}, \bibinfo {author} {\bibfnamefont {M.~E.}\ \bibnamefont
  {Schwartz}}, \bibinfo {author} {\bibfnamefont {J.}~\bibnamefont
  {Braumüller}}, \bibinfo {author} {\bibfnamefont {P.}~\bibnamefont {Krantz}},
  \bibinfo {author} {\bibfnamefont {J.~I.-J.}\ \bibnamefont {Wang}}, \bibinfo
  {author} {\bibfnamefont {S.}~\bibnamefont {Gustavsson}},\ and\ \bibinfo
  {author} {\bibfnamefont {W.~D.}\ \bibnamefont {Oliver}},\ }\bibfield
  {booktitle} {\emph {\bibinfo {booktitle} {ANNUAL REVIEW OF CONDENSED MATTER
  PHYSICS, VOL 11, 2020}},\ }\href
  {https://doi.org/10.1146/annurev-conmatphys-031119-050605} {\bibfield
  {journal} {\bibinfo  {journal} {Annual Review of Condensed Matter Physics}\
  }\bibinfo {series} {Annual Review of Condensed Matter Physics},\ \textbf
  {\bibinfo {volume} {11}},\ \bibinfo {pages} {369} (\bibinfo {year} {2020})},\
  \Eprint
  {https://arxiv.org/abs/https://doi.org/10.1146/annurev-conmatphys-031119-050605}
  {https://doi.org/10.1146/annurev-conmatphys-031119-050605} \BibitemShut
  {NoStop}%
\bibitem [{\citenamefont {Huang}\ \emph {et~al.}(2020)\citenamefont {Huang},
  \citenamefont {Wu}, \citenamefont {Fan},\ and\ \citenamefont
  {Zhu}}]{Huang2020}%
  \BibitemOpen
  \bibfield  {author} {\bibinfo {author} {\bibfnamefont {H.-L.}\ \bibnamefont
  {Huang}}, \bibinfo {author} {\bibfnamefont {D.}~\bibnamefont {Wu}}, \bibinfo
  {author} {\bibfnamefont {D.}~\bibnamefont {Fan}},\ and\ \bibinfo {author}
  {\bibfnamefont {X.}~\bibnamefont {Zhu}},\ }\bibfield  {journal} {\bibinfo
  {journal} {SCIENCE CHINA-INFORMATION SCIENCES}\ }\textbf {\bibinfo {volume}
  {63}},\ \href {https://doi.org/10.1007/s11432-020-2881-9}
  {10.1007/s11432-020-2881-9} (\bibinfo {year} {2020})\BibitemShut {NoStop}%
\bibitem [{\citenamefont {He}\ \emph {et~al.}(2021)\citenamefont {He},
  \citenamefont {Geng}, \citenamefont {Huang}, \citenamefont {Liu},\ and\
  \citenamefont {Chen}}]{He2021}%
  \BibitemOpen
  \bibfield  {author} {\bibinfo {author} {\bibfnamefont {K.}~\bibnamefont
  {He}}, \bibinfo {author} {\bibfnamefont {X.}~\bibnamefont {Geng}}, \bibinfo
  {author} {\bibfnamefont {R.}~\bibnamefont {Huang}}, \bibinfo {author}
  {\bibfnamefont {J.}~\bibnamefont {Liu}},\ and\ \bibinfo {author}
  {\bibfnamefont {W.}~\bibnamefont {Chen}},\ }\bibfield  {journal} {\bibinfo
  {journal} {CHINESE PHYSICS B}\ }\textbf {\bibinfo {volume} {30}},\ \href
  {https://doi.org/10.1088/1674-1056/ac16cf} {10.1088/1674-1056/ac16cf}
  (\bibinfo {year} {2021})\BibitemShut {NoStop}%
\bibitem [{\citenamefont {Siddiqi}(2021)}]{Siddiqi2021}%
  \BibitemOpen
  \bibfield  {author} {\bibinfo {author} {\bibfnamefont {I.}~\bibnamefont
  {Siddiqi}},\ }\href {https://doi.org/10.1038/s41578-021-00370-4} {\bibfield
  {journal} {\bibinfo  {journal} {NATURE REVIEWS MATERIALS}\ }\textbf {\bibinfo
  {volume} {6}},\ \bibinfo {pages} {875} (\bibinfo {year} {2021})}\BibitemShut
  {NoStop}%
\bibitem [{\citenamefont {Xiong}\ \emph {et~al.}(2022)\citenamefont {Xiong},
  \citenamefont {Feng}, \citenamefont {Zheng}, \citenamefont {Cui},
  \citenamefont {Yung}, \citenamefont {Zhang}, \citenamefont {Li},\ and\
  \citenamefont {Yang}}]{Xiong2022}%
  \BibitemOpen
  \bibfield  {author} {\bibinfo {author} {\bibfnamefont {K.}~\bibnamefont
  {Xiong}}, \bibinfo {author} {\bibfnamefont {J.}~\bibnamefont {Feng}},
  \bibinfo {author} {\bibfnamefont {Y.}~\bibnamefont {Zheng}}, \bibinfo
  {author} {\bibfnamefont {J.}~\bibnamefont {Cui}}, \bibinfo {author}
  {\bibfnamefont {M.}~\bibnamefont {Yung}}, \bibinfo {author} {\bibfnamefont
  {S.}~\bibnamefont {Zhang}}, \bibinfo {author} {\bibfnamefont
  {S.}~\bibnamefont {Li}},\ and\ \bibinfo {author} {\bibfnamefont
  {H.}~\bibnamefont {Yang}},\ }\href {https://doi.org/10.1360/TB-2021-0479}
  {\bibfield  {journal} {\bibinfo  {journal} {CHINESE SCIENCE
  BULLETIN-CHINESE}\ }\textbf {\bibinfo {volume} {67}},\ \bibinfo {pages} {143}
  (\bibinfo {year} {2022})}\BibitemShut {NoStop}%
\bibitem [{\citenamefont {Ezratty}(2023)}]{Ezratty2023}%
  \BibitemOpen
  \bibfield  {author} {\bibinfo {author} {\bibfnamefont {O.}~\bibnamefont
  {Ezratty}},\ }\bibfield  {journal} {\bibinfo  {journal} {EUROPEAN PHYSICAL
  JOURNAL A}\ }\textbf {\bibinfo {volume} {59}},\ \href
  {https://doi.org/10.1140/epja/s10050-023-01006-7}
  {10.1140/epja/s10050-023-01006-7} (\bibinfo {year} {2023})\BibitemShut
  {NoStop}%
\bibitem [{\citenamefont {Padamsee}(2009)}]{padamsee2009}%
  \BibitemOpen
  \bibfield  {author} {\bibinfo {author} {\bibfnamefont {H.}~\bibnamefont
  {Padamsee}},\ }\href {https://books.google.com/books?id=vFE0LVUZmgUC} {\emph
  {\bibinfo {title} {RF Superconductivity: Science, Technology, and
  Applications}}},\ Rf Superconductivity\ (\bibinfo  {publisher} {Wiley},\
  \bibinfo {year} {2009})\BibitemShut {NoStop}%
\bibitem [{\citenamefont {Valente-Feliciano}\ \emph {et~al.}(2022)\citenamefont
  {Valente-Feliciano}, \citenamefont {Antoine}, \citenamefont {Anlage},
  \citenamefont {Ciovati}, \citenamefont {Delayen}, \citenamefont {Gerigk},
  \citenamefont {Gurevich}, \citenamefont {Junginger}, \citenamefont {Keckert},
  \citenamefont {Keppe}, \citenamefont {Knobloch}, \citenamefont {Kubo},
  \citenamefont {Kugeler}, \citenamefont {Manos}, \citenamefont {Pira},
  \citenamefont {Proslier}, \citenamefont {Pudasaini}, \citenamefont {Reece},
  \citenamefont {Rimmer},\ and\ \citenamefont {Wenskat}}]{Anne_2022}%
  \BibitemOpen
  \bibfield  {author} {\bibinfo {author} {\bibfnamefont {A.-M.}\ \bibnamefont
  {Valente-Feliciano}}, \bibinfo {author} {\bibfnamefont {C.}~\bibnamefont
  {Antoine}}, \bibinfo {author} {\bibfnamefont {S.}~\bibnamefont {Anlage}},
  \bibinfo {author} {\bibfnamefont {G.}~\bibnamefont {Ciovati}}, \bibinfo
  {author} {\bibfnamefont {J.}~\bibnamefont {Delayen}}, \bibinfo {author}
  {\bibfnamefont {F.}~\bibnamefont {Gerigk}}, \bibinfo {author} {\bibfnamefont
  {A.}~\bibnamefont {Gurevich}}, \bibinfo {author} {\bibfnamefont
  {T.}~\bibnamefont {Junginger}}, \bibinfo {author} {\bibfnamefont
  {S.}~\bibnamefont {Keckert}}, \bibinfo {author} {\bibfnamefont
  {G.}~\bibnamefont {Keppe}}, \bibinfo {author} {\bibfnamefont
  {J.}~\bibnamefont {Knobloch}}, \bibinfo {author} {\bibfnamefont
  {T.}~\bibnamefont {Kubo}}, \bibinfo {author} {\bibfnamefont {O.}~\bibnamefont
  {Kugeler}}, \bibinfo {author} {\bibfnamefont {D.}~\bibnamefont {Manos}},
  \bibinfo {author} {\bibfnamefont {C.}~\bibnamefont {Pira}}, \bibinfo {author}
  {\bibfnamefont {T.}~\bibnamefont {Proslier}}, \bibinfo {author}
  {\bibfnamefont {U.}~\bibnamefont {Pudasaini}}, \bibinfo {author}
  {\bibfnamefont {C.}~\bibnamefont {Reece}}, \bibinfo {author} {\bibfnamefont
  {R.}~\bibnamefont {Rimmer}},\ and\ \bibinfo {author} {\bibfnamefont
  {M.}~\bibnamefont {Wenskat}},\ }\href@noop {} {\bibfield  {journal} {\bibinfo
   {journal} {arXiv}\ } (\bibinfo {year} {2022})}\BibitemShut {NoStop}%
\bibitem [{\citenamefont {Stromberg}(1965)}]{Stromberg1965}%
  \BibitemOpen
  \bibfield  {author} {\bibinfo {author} {\bibfnamefont {T.~F.}\ \bibnamefont
  {Stromberg}},\ }\emph {\bibinfo {title} {The superconducting properties of
  high purity niobium}},\ \href
  {https://dr.lib.iastate.edu/entities/publication/0ea2ad68-bcef-4712-b42d-b466d8de33cb}
  {Ph.D. thesis},\ \bibinfo  {school} {Iowa State University} (\bibinfo {year}
  {1965})\BibitemShut {NoStop}%
\bibitem [{\citenamefont {Finnemore}\ \emph {et~al.}(1966)\citenamefont
  {Finnemore}, \citenamefont {Stromberg},\ and\ \citenamefont
  {Swenson}}]{Finnemore1966}%
  \BibitemOpen
  \bibfield  {author} {\bibinfo {author} {\bibfnamefont {D.~K.}\ \bibnamefont
  {Finnemore}}, \bibinfo {author} {\bibfnamefont {T.~F.}\ \bibnamefont
  {Stromberg}},\ and\ \bibinfo {author} {\bibfnamefont {C.~A.}\ \bibnamefont
  {Swenson}},\ }\href {https://doi.org/10.1103/physrev.149.231} {\bibfield
  {journal} {\bibinfo  {journal} {Physical Review}\ }\textbf {\bibinfo {volume}
  {149}},\ \bibinfo {pages} {231} (\bibinfo {year} {1966})}\BibitemShut
  {NoStop}%
\bibitem [{\citenamefont {Daams}\ and\ \citenamefont
  {Carbotte}(1980)}]{Daams1980}%
  \BibitemOpen
  \bibfield  {author} {\bibinfo {author} {\bibfnamefont {J.}~\bibnamefont
  {Daams}}\ and\ \bibinfo {author} {\bibfnamefont {J.~P.}\ \bibnamefont
  {Carbotte}},\ }\href {https://doi.org/10.1007/BF00115987} {\bibfield
  {journal} {\bibinfo  {journal} {J. Low Temp. Phys.}\ }\textbf {\bibinfo
  {volume} {40}},\ \bibinfo {pages} {135} (\bibinfo {year} {1980})}\BibitemShut
  {NoStop}%
\bibitem [{\citenamefont {Bahte}\ \emph {et~al.}(1998)\citenamefont {Bahte},
  \citenamefont {Herrmann},\ and\ \citenamefont {Schmuser}}]{Bahte1998}%
  \BibitemOpen
  \bibfield  {author} {\bibinfo {author} {\bibfnamefont {M.}~\bibnamefont
  {Bahte}}, \bibinfo {author} {\bibfnamefont {F.}~\bibnamefont {Herrmann}},\
  and\ \bibinfo {author} {\bibfnamefont {P.}~\bibnamefont {Schmuser}},\
  }\href@noop {} {\bibfield  {journal} {\bibinfo  {journal} {Part. Accel.}\
  }\textbf {\bibinfo {volume} {60}},\ \bibinfo {pages} {121} (\bibinfo {year}
  {1998})}\BibitemShut {NoStop}%
\bibitem [{\citenamefont {Koethe}\ and\ \citenamefont
  {Moench}(2000)}]{Koethe2000}%
  \BibitemOpen
  \bibfield  {author} {\bibinfo {author} {\bibfnamefont {A.}~\bibnamefont
  {Koethe}}\ and\ \bibinfo {author} {\bibfnamefont {J.~I.}\ \bibnamefont
  {Moench}},\ }\href {https://doi.org/10.2320/matertrans1989.41.7} {\bibfield
  {journal} {\bibinfo  {journal} {Materials Transactions, JIM}\ }\textbf
  {\bibinfo {volume} {41}},\ \bibinfo {pages} {7} (\bibinfo {year}
  {2000})}\BibitemShut {NoStop}%
\bibitem [{\citenamefont {Prozorov}\ \emph {et~al.}(2006)\citenamefont
  {Prozorov}, \citenamefont {Shantsev},\ and\ \citenamefont
  {Mints}}]{Prozorov2006a}%
  \BibitemOpen
  \bibfield  {author} {\bibinfo {author} {\bibfnamefont {R.}~\bibnamefont
  {Prozorov}}, \bibinfo {author} {\bibfnamefont {D.~V.}\ \bibnamefont
  {Shantsev}},\ and\ \bibinfo {author} {\bibfnamefont {R.~G.}\ \bibnamefont
  {Mints}},\ }\href {https://doi.org/10.1103/PhysRevB.74.220511} {\bibfield
  {journal} {\bibinfo  {journal} {Phys. Rev. B}\ }\textbf {\bibinfo {volume}
  {74}},\ \bibinfo {pages} {220511} (\bibinfo {year} {2006})}\BibitemShut
  {NoStop}%
\bibitem [{\citenamefont {Kozhevnikov}\ \emph {et~al.}(2017)\citenamefont
  {Kozhevnikov}, \citenamefont {Valente-Feliciano}, \citenamefont {Curran},
  \citenamefont {Suter}, \citenamefont {Liu}, \citenamefont {Richter},
  \citenamefont {Morenzoni}, \citenamefont {Bending},\ and\ \citenamefont
  {Haesendonck}}]{Kozhevnikov2017}%
  \BibitemOpen
  \bibfield  {author} {\bibinfo {author} {\bibfnamefont {V.}~\bibnamefont
  {Kozhevnikov}}, \bibinfo {author} {\bibfnamefont {A.-M.}\ \bibnamefont
  {Valente-Feliciano}}, \bibinfo {author} {\bibfnamefont {P.~J.}\ \bibnamefont
  {Curran}}, \bibinfo {author} {\bibfnamefont {A.}~\bibnamefont {Suter}},
  \bibinfo {author} {\bibfnamefont {A.~H.}\ \bibnamefont {Liu}}, \bibinfo
  {author} {\bibfnamefont {G.}~\bibnamefont {Richter}}, \bibinfo {author}
  {\bibfnamefont {E.}~\bibnamefont {Morenzoni}}, \bibinfo {author}
  {\bibfnamefont {S.~J.}\ \bibnamefont {Bending}},\ and\ \bibinfo {author}
  {\bibfnamefont {C.~V.}\ \bibnamefont {Haesendonck}},\ }\href
  {https://doi.org/10.1103/physrevb.95.174509} {\bibfield  {journal} {\bibinfo
  {journal} {Physical Review B}\ }\textbf {\bibinfo {volume} {95}},\ \bibinfo
  {pages} {174509} (\bibinfo {year} {2017})}\BibitemShut {NoStop}%
\bibitem [{\citenamefont {Liarte}\ \emph {et~al.}(2017)\citenamefont {Liarte},
  \citenamefont {Posen}, \citenamefont {Transtrum}, \citenamefont {Catelani},
  \citenamefont {Liepe},\ and\ \citenamefont {Sethna}}]{Liarte_2017}%
  \BibitemOpen
  \bibfield  {author} {\bibinfo {author} {\bibfnamefont {D.~B.}\ \bibnamefont
  {Liarte}}, \bibinfo {author} {\bibfnamefont {S.}~\bibnamefont {Posen}},
  \bibinfo {author} {\bibfnamefont {M.~K.}\ \bibnamefont {Transtrum}}, \bibinfo
  {author} {\bibfnamefont {G.}~\bibnamefont {Catelani}}, \bibinfo {author}
  {\bibfnamefont {M.}~\bibnamefont {Liepe}},\ and\ \bibinfo {author}
  {\bibfnamefont {J.~P.}\ \bibnamefont {Sethna}},\ }\href
  {https://doi.org/10.1088/1361-6668/30/3/033002} {\bibfield  {journal}
  {\bibinfo  {journal} {Superconductor Science and Technology}\ }\textbf
  {\bibinfo {volume} {30}},\ \bibinfo {pages} {033002} (\bibinfo {year}
  {2017})}\BibitemShut {NoStop}%
\bibitem [{\citenamefont {Wendin}(2017)}]{wendin2017quantum}%
  \BibitemOpen
  \bibfield  {author} {\bibinfo {author} {\bibfnamefont {G.}~\bibnamefont
  {Wendin}},\ }\href@noop {} {\bibfield  {journal} {\bibinfo  {journal}
  {Reports on Progress in Physics}\ }\textbf {\bibinfo {volume} {80}},\
  \bibinfo {pages} {106001} (\bibinfo {year} {2017})}\BibitemShut {NoStop}%
\bibitem [{\citenamefont {Padamsee}\ \emph {et~al.}(2008)\citenamefont
  {Padamsee}, \citenamefont {Knobloch},\ and\ \citenamefont
  {Hays}}]{padamsee2008}%
  \BibitemOpen
  \bibfield  {author} {\bibinfo {author} {\bibfnamefont {H.}~\bibnamefont
  {Padamsee}}, \bibinfo {author} {\bibfnamefont {J.}~\bibnamefont {Knobloch}},\
  and\ \bibinfo {author} {\bibfnamefont {T.}~\bibnamefont {Hays}},\ }\href
  {https://books.google.com/books?id=VFG7EAAAQBAJ} {\emph {\bibinfo {title} {RF
  Superconductivity for Accelerators}}}\ (\bibinfo  {publisher} {Wiley},\
  \bibinfo {year} {2008})\BibitemShut {NoStop}%
\bibitem [{\citenamefont {Gurevich}(2012)}]{Gurevich2012}%
  \BibitemOpen
  \bibfield  {author} {\bibinfo {author} {\bibfnamefont {A.}~\bibnamefont
  {Gurevich}},\ }\href {https://doi.org/10.1142/s1793626812300058} {\bibfield
  {journal} {\bibinfo  {journal} {Reviews of Accelerator Science and
  Technology}\ }\textbf {\bibinfo {volume} {05}},\ \bibinfo {pages} {119}
  (\bibinfo {year} {2012})}\BibitemShut {NoStop}%
\bibitem [{\citenamefont {Gurevich}(2017)}]{Gurevich2017a}%
  \BibitemOpen
  \bibfield  {author} {\bibinfo {author} {\bibfnamefont {A.}~\bibnamefont
  {Gurevich}},\ }\href {https://doi.org/10.1088/1361-6668/30/3/034004}
  {\bibfield  {journal} {\bibinfo  {journal} {Superconductor Science and
  Technology}\ }\textbf {\bibinfo {volume} {30}},\ \bibinfo {pages} {034004}
  (\bibinfo {year} {2017})}\BibitemShut {NoStop}%
\bibitem [{\citenamefont {Ngampruetikorn}\ and\ \citenamefont
  {Sauls}(2019)}]{Ngampruetikorn2019}%
  \BibitemOpen
  \bibfield  {author} {\bibinfo {author} {\bibfnamefont {V.}~\bibnamefont
  {Ngampruetikorn}}\ and\ \bibinfo {author} {\bibfnamefont {J.~A.}\
  \bibnamefont {Sauls}},\ }\href
  {https://doi.org/10.1103/physrevresearch.1.012015} {\bibfield  {journal}
  {\bibinfo  {journal} {Physical Review Research}\ }\textbf {\bibinfo {volume}
  {1}},\ \bibinfo {pages} {012015} (\bibinfo {year} {2019})}\BibitemShut
  {NoStop}%
\bibitem [{\citenamefont {Ueki}\ \emph
  {et~al.}(2022{\natexlab{a}})\citenamefont {Ueki}, \citenamefont {Zarea},\
  and\ \citenamefont {Sauls}}]{Ueki2022}%
  \BibitemOpen
  \bibfield  {author} {\bibinfo {author} {\bibfnamefont {H.}~\bibnamefont
  {Ueki}}, \bibinfo {author} {\bibfnamefont {M.}~\bibnamefont {Zarea}},\ and\
  \bibinfo {author} {\bibfnamefont {J.~A.}\ \bibnamefont {Sauls}}\ }\href
  {https://doi.org/10.48550/ARXIV.2207.14236} {10.48550/ARXIV.2207.14236}
  (\bibinfo {year} {2022}{\natexlab{a}}),\ \Eprint
  {https://arxiv.org/abs/2207.14236} {arXiv:2207.14236 [cond-mat.supr-con]}
  \BibitemShut {NoStop}%
\bibitem [{\citenamefont {Ueki}\ \emph
  {et~al.}(2022{\natexlab{b}})\citenamefont {Ueki}, \citenamefont {Zarea},\
  and\ \citenamefont {Sauls}}]{Ueki2022a}%
  \BibitemOpen
  \bibfield  {author} {\bibinfo {author} {\bibfnamefont {H.}~\bibnamefont
  {Ueki}}, \bibinfo {author} {\bibfnamefont {M.}~\bibnamefont {Zarea}},\ and\
  \bibinfo {author} {\bibfnamefont {J.~A.}\ \bibnamefont {Sauls}}\ }\href
  {https://doi.org/10.48550/ARXIV.2209.11752} {10.48550/ARXIV.2209.11752}
  (\bibinfo {year} {2022}{\natexlab{b}})\BibitemShut {NoStop}%
\bibitem [{\citenamefont {Daunt}\ \emph {et~al.}(1937)\citenamefont {Daunt},
  \citenamefont {Mendelssohn},\ and\ \citenamefont {Lindemann}}]{Daunt1937}%
  \BibitemOpen
  \bibfield  {author} {\bibinfo {author} {\bibfnamefont {J.~G.}\ \bibnamefont
  {Daunt}}, \bibinfo {author} {\bibfnamefont {K.}~\bibnamefont {Mendelssohn}},\
  and\ \bibinfo {author} {\bibfnamefont {F.~A.}\ \bibnamefont {Lindemann}},\
  }\href {https://doi.org/10.1098/rspa.1937.0099} {\bibfield  {journal}
  {\bibinfo  {journal} {Proc. Roy. Soc. London. Ser. A - Math. Phys. Sci.}\
  }\textbf {\bibinfo {volume} {160}},\ \bibinfo {pages} {127} (\bibinfo {year}
  {1937})}\BibitemShut {NoStop}%
\bibitem [{\citenamefont {Zarea}\ \emph {et~al.}(2022)\citenamefont {Zarea},
  \citenamefont {Ueki},\ and\ \citenamefont {Sauls}}]{Zarea2022}%
  \BibitemOpen
  \bibfield  {author} {\bibinfo {author} {\bibfnamefont {M.}~\bibnamefont
  {Zarea}}, \bibinfo {author} {\bibfnamefont {H.}~\bibnamefont {Ueki}},\ and\
  \bibinfo {author} {\bibfnamefont {J.~A.}\ \bibnamefont {Sauls}},\ }\href@noop
  {} {\bibfield  {journal} {\bibinfo  {journal} {arXiv:2201.07403}\ } (\bibinfo
  {year} {2022})},\ \Eprint {https://arxiv.org/abs/2201.07403}
  {arXiv:2201.07403 [cond-mat.supr-con]} \BibitemShut {NoStop}%
\bibitem [{\citenamefont {Prozorov}\ \emph {et~al.}(2022)\citenamefont
  {Prozorov}, \citenamefont {Zarea},\ and\ \citenamefont
  {Sauls}}]{Prozorov2022}%
  \BibitemOpen
  \bibfield  {author} {\bibinfo {author} {\bibfnamefont {R.}~\bibnamefont
  {Prozorov}}, \bibinfo {author} {\bibfnamefont {M.}~\bibnamefont {Zarea}},\
  and\ \bibinfo {author} {\bibfnamefont {J.~A.}\ \bibnamefont {Sauls}},\ }\href
  {https://doi.org/10.1103/PhysRevB.106.L180505} {\bibfield  {journal}
  {\bibinfo  {journal} {Phys. Rev. B}\ }\textbf {\bibinfo {volume} {106}},\
  \bibinfo {pages} {L180505} (\bibinfo {year} {2022})}\BibitemShut {NoStop}%
\bibitem [{\citenamefont {Barkov}\ \emph {et~al.}(2012)\citenamefont {Barkov},
  \citenamefont {Romanenko},\ and\ \citenamefont {Grassellino}}]{Barkov2012}%
  \BibitemOpen
  \bibfield  {author} {\bibinfo {author} {\bibfnamefont {F.}~\bibnamefont
  {Barkov}}, \bibinfo {author} {\bibfnamefont {A.}~\bibnamefont {Romanenko}},\
  and\ \bibinfo {author} {\bibfnamefont {A.}~\bibnamefont {Grassellino}},\
  }\href {https://doi.org/10.1103/physrevstab.15.122001} {\bibfield  {journal}
  {\bibinfo  {journal} {Physical Review Special Topics - Accelerators and
  Beams}\ }\textbf {\bibinfo {volume} {15}},\ \bibinfo {pages} {122001}
  (\bibinfo {year} {2012})}\BibitemShut {NoStop}%
\bibitem [{\citenamefont {Barkov}\ \emph {et~al.}(2013)\citenamefont {Barkov},
  \citenamefont {Romanenko}, \citenamefont {Trenikhina},\ and\ \citenamefont
  {Grassellino}}]{Barkov2013}%
  \BibitemOpen
  \bibfield  {author} {\bibinfo {author} {\bibfnamefont {F.}~\bibnamefont
  {Barkov}}, \bibinfo {author} {\bibfnamefont {A.}~\bibnamefont {Romanenko}},
  \bibinfo {author} {\bibfnamefont {Y.}~\bibnamefont {Trenikhina}},\ and\
  \bibinfo {author} {\bibfnamefont {A.}~\bibnamefont {Grassellino}},\ }\href
  {https://doi.org/10.1063/1.4826901} {\bibfield  {journal} {\bibinfo
  {journal} {Journal of Applied Physics}\ }\textbf {\bibinfo {volume} {114}},\
  \bibinfo {pages} {164904} (\bibinfo {year} {2013})}\BibitemShut {NoStop}%
\bibitem [{\citenamefont {Romanenko}\ \emph {et~al.}(2020)\citenamefont
  {Romanenko}, \citenamefont {Pilipenko}, \citenamefont {Zorzetti},
  \citenamefont {Frolov}, \citenamefont {Awida}, \citenamefont {Belomestnykh},
  \citenamefont {Posen},\ and\ \citenamefont {Grassellino}}]{alex20}%
  \BibitemOpen
  \bibfield  {author} {\bibinfo {author} {\bibfnamefont {A.}~\bibnamefont
  {Romanenko}}, \bibinfo {author} {\bibfnamefont {R.}~\bibnamefont
  {Pilipenko}}, \bibinfo {author} {\bibfnamefont {S.}~\bibnamefont {Zorzetti}},
  \bibinfo {author} {\bibfnamefont {D.}~\bibnamefont {Frolov}}, \bibinfo
  {author} {\bibfnamefont {M.}~\bibnamefont {Awida}}, \bibinfo {author}
  {\bibfnamefont {S.}~\bibnamefont {Belomestnykh}}, \bibinfo {author}
  {\bibfnamefont {S.}~\bibnamefont {Posen}},\ and\ \bibinfo {author}
  {\bibfnamefont {A.}~\bibnamefont {Grassellino}},\ }\href
  {https://doi.org/10.1103/PhysRevApplied.13.034032} {\bibfield  {journal}
  {\bibinfo  {journal} {Phys. Rev. Appl.}\ }\textbf {\bibinfo {volume} {13}},\
  \bibinfo {pages} {034032} (\bibinfo {year} {2020})}\BibitemShut {NoStop}%
\bibitem [{\citenamefont {Burnett}\ \emph {et~al.}(2016)\citenamefont
  {Burnett}, \citenamefont {Faoro},\ and\ \citenamefont
  {Lindstrom}}]{Burnett2016}%
  \BibitemOpen
  \bibfield  {author} {\bibinfo {author} {\bibfnamefont {J.}~\bibnamefont
  {Burnett}}, \bibinfo {author} {\bibfnamefont {L.}~\bibnamefont {Faoro}},\
  and\ \bibinfo {author} {\bibfnamefont {T.}~\bibnamefont {Lindstrom}},\ }\href
  {https://doi.org/10.1088/0953-2048/29/4/044008} {\bibfield  {journal}
  {\bibinfo  {journal} {Superconductor Science and Technology}\ }\textbf
  {\bibinfo {volume} {29}},\ \bibinfo {pages} {044008} (\bibinfo {year}
  {2016})}\BibitemShut {NoStop}%
\bibitem [{\citenamefont {Müller}\ \emph {et~al.}(2019)\citenamefont
  {Müller}, \citenamefont {Cole},\ and\ \citenamefont
  {Lisenfeld}}]{Mueller2019}%
  \BibitemOpen
  \bibfield  {author} {\bibinfo {author} {\bibfnamefont {C.}~\bibnamefont
  {Müller}}, \bibinfo {author} {\bibfnamefont {J.~H.}\ \bibnamefont {Cole}},\
  and\ \bibinfo {author} {\bibfnamefont {J.}~\bibnamefont {Lisenfeld}},\ }\href
  {https://doi.org/10.1088/1361-6633/ab3a7e} {\bibfield  {journal} {\bibinfo
  {journal} {Reports on Progress in Physics}\ }\textbf {\bibinfo {volume}
  {82}},\ \bibinfo {pages} {124501} (\bibinfo {year} {2019})}\BibitemShut
  {NoStop}%
\bibitem [{\citenamefont {McRae}\ \emph {et~al.}(2020)\citenamefont {McRae},
  \citenamefont {Wang}, \citenamefont {Gao}, \citenamefont {Vissers},
  \citenamefont {Brecht}, \citenamefont {Dunsworth}, \citenamefont {Pappas},\
  and\ \citenamefont {Mutus}}]{McRae20}%
  \BibitemOpen
  \bibfield  {author} {\bibinfo {author} {\bibfnamefont {C.~R.~H.}\
  \bibnamefont {McRae}}, \bibinfo {author} {\bibfnamefont {H.}~\bibnamefont
  {Wang}}, \bibinfo {author} {\bibfnamefont {J.}~\bibnamefont {Gao}}, \bibinfo
  {author} {\bibfnamefont {M.~R.}\ \bibnamefont {Vissers}}, \bibinfo {author}
  {\bibfnamefont {T.}~\bibnamefont {Brecht}}, \bibinfo {author} {\bibfnamefont
  {A.}~\bibnamefont {Dunsworth}}, \bibinfo {author} {\bibfnamefont {D.~P.}\
  \bibnamefont {Pappas}},\ and\ \bibinfo {author} {\bibfnamefont
  {J.}~\bibnamefont {Mutus}},\ }\bibfield  {journal} {\bibinfo  {journal}
  {Review of Scientific Instruments}\ }\textbf {\bibinfo {volume} {91}},\ \href
  {https://doi.org/10.1063/5.0017378} {10.1063/5.0017378} (\bibinfo {year}
  {2020}),\ \bibinfo {note} {091101},\ \Eprint
  {https://arxiv.org/abs/https://pubs.aip.org/aip/rsi/article-pdf/doi/10.1063/5.0017378/14797873/091101\_1\_online.pdf}
  {https://pubs.aip.org/aip/rsi/article-pdf/doi/10.1063/5.0017378/14797873/091101\_1\_online.pdf}
  \BibitemShut {NoStop}%
\bibitem [{\citenamefont {Preskill}(2018)}]{preskill2018quantum}%
  \BibitemOpen
  \bibfield  {author} {\bibinfo {author} {\bibfnamefont {J.}~\bibnamefont
  {Preskill}},\ }\href@noop {} {\emph {\bibinfo {title} {Quantum computing in
  the NISQ era and beyond}}}\ (\bibinfo  {publisher} {Quantum},\ \bibinfo
  {year} {2018})\BibitemShut {NoStop}%
\bibitem [{\citenamefont {An}\ \emph {et~al.}(2003)\citenamefont {An},
  \citenamefont {Fukuyama}, \citenamefont {Yokogawa},\ and\ \citenamefont
  {Yoshimura}}]{An_oxide_2003}%
  \BibitemOpen
  \bibfield  {author} {\bibinfo {author} {\bibfnamefont {B.}~\bibnamefont
  {An}}, \bibinfo {author} {\bibfnamefont {S.}~\bibnamefont {Fukuyama}},
  \bibinfo {author} {\bibfnamefont {K.}~\bibnamefont {Yokogawa}},\ and\
  \bibinfo {author} {\bibfnamefont {M.}~\bibnamefont {Yoshimura}},\ }\href
  {https://doi.org/10.1103/PhysRevB.68.115423} {\bibfield  {journal} {\bibinfo
  {journal} {Phys. Rev. B}\ }\textbf {\bibinfo {volume} {68}},\ \bibinfo
  {pages} {115423} (\bibinfo {year} {2003})}\BibitemShut {NoStop}%
\bibitem [{\citenamefont {Samsonova}\ \emph {et~al.}(2021)\citenamefont
  {Samsonova}, \citenamefont {Zolotov}, \citenamefont {Baeva}, \citenamefont
  {Lomakin}, \citenamefont {Titova}, \citenamefont {Kardakova},\ and\
  \citenamefont {Goltsman}}]{Samsonova2021}%
  \BibitemOpen
  \bibfield  {author} {\bibinfo {author} {\bibfnamefont {A.~S.}\ \bibnamefont
  {Samsonova}}, \bibinfo {author} {\bibfnamefont {P.~I.}\ \bibnamefont
  {Zolotov}}, \bibinfo {author} {\bibfnamefont {E.~M.}\ \bibnamefont {Baeva}},
  \bibinfo {author} {\bibfnamefont {A.~I.}\ \bibnamefont {Lomakin}}, \bibinfo
  {author} {\bibfnamefont {N.~A.}\ \bibnamefont {Titova}}, \bibinfo {author}
  {\bibfnamefont {A.~I.}\ \bibnamefont {Kardakova}},\ and\ \bibinfo {author}
  {\bibfnamefont {G.~N.}\ \bibnamefont {Goltsman}},\ }\bibfield  {journal}
  {\bibinfo  {journal} {IEEE Trans. Appl. Supercond.}\ }\textbf {\bibinfo
  {volume} {31}},\ \href {https://doi.org/10.1109/TASC.2021.3065281}
  {10.1109/TASC.2021.3065281} (\bibinfo {year} {2021})\BibitemShut {NoStop}%
\bibitem [{\citenamefont {Verjauw}\ \emph {et~al.}(2021)\citenamefont
  {Verjauw}, \citenamefont {Poto\ifmmode~\check{c}\else \v{c}\fi{}nik},
  \citenamefont {Mongillo}, \citenamefont {Acharya}, \citenamefont
  {Mohiyaddin}, \citenamefont {Simion}, \citenamefont {Pacco}, \citenamefont
  {Ivanov}, \citenamefont {Wan}, \citenamefont {Vanleenhove}, \citenamefont
  {Souriau}, \citenamefont {Jussot}, \citenamefont {Thiam}, \citenamefont
  {Swerts}, \citenamefont {Piao}, \citenamefont {Couet}, \citenamefont {Heyns},
  \citenamefont {Govoreanu},\ and\ \citenamefont {Radu}}]{Verjauw2021}%
  \BibitemOpen
  \bibfield  {author} {\bibinfo {author} {\bibfnamefont {J.}~\bibnamefont
  {Verjauw}}, \bibinfo {author} {\bibfnamefont {A.}~\bibnamefont
  {Poto\ifmmode~\check{c}\else \v{c}\fi{}nik}}, \bibinfo {author}
  {\bibfnamefont {M.}~\bibnamefont {Mongillo}}, \bibinfo {author}
  {\bibfnamefont {R.}~\bibnamefont {Acharya}}, \bibinfo {author} {\bibfnamefont
  {F.}~\bibnamefont {Mohiyaddin}}, \bibinfo {author} {\bibfnamefont
  {G.}~\bibnamefont {Simion}}, \bibinfo {author} {\bibfnamefont
  {A.}~\bibnamefont {Pacco}}, \bibinfo {author} {\bibfnamefont
  {T.}~\bibnamefont {Ivanov}}, \bibinfo {author} {\bibfnamefont
  {D.}~\bibnamefont {Wan}}, \bibinfo {author} {\bibfnamefont {A.}~\bibnamefont
  {Vanleenhove}}, \bibinfo {author} {\bibfnamefont {L.}~\bibnamefont
  {Souriau}}, \bibinfo {author} {\bibfnamefont {J.}~\bibnamefont {Jussot}},
  \bibinfo {author} {\bibfnamefont {A.}~\bibnamefont {Thiam}}, \bibinfo
  {author} {\bibfnamefont {J.}~\bibnamefont {Swerts}}, \bibinfo {author}
  {\bibfnamefont {X.}~\bibnamefont {Piao}}, \bibinfo {author} {\bibfnamefont
  {S.}~\bibnamefont {Couet}}, \bibinfo {author} {\bibfnamefont
  {M.}~\bibnamefont {Heyns}}, \bibinfo {author} {\bibfnamefont
  {B.}~\bibnamefont {Govoreanu}},\ and\ \bibinfo {author} {\bibfnamefont
  {I.}~\bibnamefont {Radu}},\ }\href
  {https://doi.org/10.1103/PhysRevApplied.16.014018} {\bibfield  {journal}
  {\bibinfo  {journal} {Phys. Rev. Applied}\ }\textbf {\bibinfo {volume}
  {16}},\ \bibinfo {pages} {014018} (\bibinfo {year} {2021})}\BibitemShut
  {NoStop}%
\bibitem [{\citenamefont {Murthy}\ \emph {et~al.}(2022)\citenamefont {Murthy},
  \citenamefont {Masih~Das}, \citenamefont {Ribet}, \citenamefont {Kopas},
  \citenamefont {Lee}, \citenamefont {Reagor}, \citenamefont {Zhou},
  \citenamefont {Kramer}, \citenamefont {Hersam}, \citenamefont {Checchin},
  \citenamefont {Grassellino}, \citenamefont {Reis}, \citenamefont {Dravid},\
  and\ \citenamefont {Romanenko}}]{Murthy22a}%
  \BibitemOpen
  \bibfield  {author} {\bibinfo {author} {\bibfnamefont {A.~A.}\ \bibnamefont
  {Murthy}}, \bibinfo {author} {\bibfnamefont {P.}~\bibnamefont {Masih~Das}},
  \bibinfo {author} {\bibfnamefont {S.~M.}\ \bibnamefont {Ribet}}, \bibinfo
  {author} {\bibfnamefont {C.}~\bibnamefont {Kopas}}, \bibinfo {author}
  {\bibfnamefont {J.}~\bibnamefont {Lee}}, \bibinfo {author} {\bibfnamefont
  {M.~J.}\ \bibnamefont {Reagor}}, \bibinfo {author} {\bibfnamefont
  {L.}~\bibnamefont {Zhou}}, \bibinfo {author} {\bibfnamefont {M.~J.}\
  \bibnamefont {Kramer}}, \bibinfo {author} {\bibfnamefont {M.~C.}\
  \bibnamefont {Hersam}}, \bibinfo {author} {\bibfnamefont {M.}~\bibnamefont
  {Checchin}}, \bibinfo {author} {\bibfnamefont {A.}~\bibnamefont
  {Grassellino}}, \bibinfo {author} {\bibfnamefont {R.~d.}\ \bibnamefont
  {Reis}}, \bibinfo {author} {\bibfnamefont {V.~P.}\ \bibnamefont {Dravid}},\
  and\ \bibinfo {author} {\bibfnamefont {A.}~\bibnamefont {Romanenko}},\ }\href
  {https://doi.org/10.1021/acsnano.2c07913} {\bibfield  {journal} {\bibinfo
  {journal} {ACS Nano}\ }\textbf {\bibinfo {volume} {16}},\ \bibinfo {pages}
  {17257} (\bibinfo {year} {2022})}\BibitemShut {NoStop}%
\bibitem [{\citenamefont {Nersisyan}\ \emph {et~al.}(2019)\citenamefont
  {Nersisyan}, \citenamefont {Poletto}, \citenamefont {Alidoust}, \citenamefont
  {Manenti}, \citenamefont {Renzas}, \citenamefont {Bui}, \citenamefont {Vu},
  \citenamefont {Whyland}, \citenamefont {Mohan}, \citenamefont {Sete},
  \citenamefont {Stanwyck}, \citenamefont {Bestwick},\ and\ \citenamefont
  {Reagor}}]{Nersisyan2019a}%
  \BibitemOpen
  \bibfield  {author} {\bibinfo {author} {\bibfnamefont {A.}~\bibnamefont
  {Nersisyan}}, \bibinfo {author} {\bibfnamefont {S.}~\bibnamefont {Poletto}},
  \bibinfo {author} {\bibfnamefont {N.}~\bibnamefont {Alidoust}}, \bibinfo
  {author} {\bibfnamefont {R.}~\bibnamefont {Manenti}}, \bibinfo {author}
  {\bibfnamefont {R.}~\bibnamefont {Renzas}}, \bibinfo {author} {\bibfnamefont
  {C.-V.}\ \bibnamefont {Bui}}, \bibinfo {author} {\bibfnamefont
  {K.}~\bibnamefont {Vu}}, \bibinfo {author} {\bibfnamefont {T.}~\bibnamefont
  {Whyland}}, \bibinfo {author} {\bibfnamefont {Y.}~\bibnamefont {Mohan}},
  \bibinfo {author} {\bibfnamefont {E.~A.}\ \bibnamefont {Sete}}, \bibinfo
  {author} {\bibfnamefont {S.}~\bibnamefont {Stanwyck}}, \bibinfo {author}
  {\bibfnamefont {A.}~\bibnamefont {Bestwick}},\ and\ \bibinfo {author}
  {\bibfnamefont {M.}~\bibnamefont {Reagor}},\ }\href@noop {} {\bibfield
  {journal} {\bibinfo  {journal} {arXiv:1901.08042}\ } (\bibinfo {year}
  {2019})},\ \Eprint {https://arxiv.org/abs/1901.08042} {arXiv:1901.08042
  [quant-ph]} \BibitemShut {NoStop}%
\bibitem [{\citenamefont {Romanenko}(2009)}]{Alex09a}%
  \BibitemOpen
  \bibfield  {author} {\bibinfo {author} {\bibfnamefont {A.}~\bibnamefont
  {Romanenko}},\ }\href@noop {} {\bibfield  {journal} {\bibinfo  {journal}
  {Ph.D. thesis, Cornell University}\ } (\bibinfo {year} {2009})}\BibitemShut
  {NoStop}%
\bibitem [{\citenamefont {Van~Degrift}(1975)}]{VanDegrift1975RSI}%
  \BibitemOpen
  \bibfield  {author} {\bibinfo {author} {\bibfnamefont {C.~T.}\ \bibnamefont
  {Van~Degrift}},\ }\href {https://doi.org/http://dx.doi.org/10.1063/1.1134272}
  {\bibfield  {journal} {\bibinfo  {journal} {Review of Scientific
  Instruments}\ }\textbf {\bibinfo {volume} {46}},\ \bibinfo {pages} {599}
  (\bibinfo {year} {1975})}\BibitemShut {NoStop}%
\bibitem [{\citenamefont {Prozorov}\ \emph
  {et~al.}(2000{\natexlab{a}})\citenamefont {Prozorov}, \citenamefont
  {Giannetta}, \citenamefont {Carrington},\ and\ \citenamefont
  {Araujo-Moreira}}]{Prozorov2000}%
  \BibitemOpen
  \bibfield  {author} {\bibinfo {author} {\bibfnamefont {R.}~\bibnamefont
  {Prozorov}}, \bibinfo {author} {\bibfnamefont {R.~W.}\ \bibnamefont
  {Giannetta}}, \bibinfo {author} {\bibfnamefont {A.}~\bibnamefont
  {Carrington}},\ and\ \bibinfo {author} {\bibfnamefont {F.~M.}\ \bibnamefont
  {Araujo-Moreira}},\ }\href {<Go to ISI>://WOS:000088037000030
  http://prb.aps.org/pdf/PRB/v62/i1/p115_1} {\bibfield  {journal} {\bibinfo
  {journal} {Phys. Rev. B}\ }\textbf {\bibinfo {volume} {62}},\ \bibinfo
  {pages} {115} (\bibinfo {year} {2000}{\natexlab{a}})}\BibitemShut {NoStop}%
\bibitem [{\citenamefont {Prozorov}\ \emph
  {et~al.}(2000{\natexlab{b}})\citenamefont {Prozorov}, \citenamefont
  {Giannetta}, \citenamefont {Carrington}, \citenamefont {Fournier},
  \citenamefont {Greene}, \citenamefont {Guptasarma}, \citenamefont {Hinks},\
  and\ \citenamefont {Banks}}]{Prozorov2000a}%
  \BibitemOpen
  \bibfield  {author} {\bibinfo {author} {\bibfnamefont {R.}~\bibnamefont
  {Prozorov}}, \bibinfo {author} {\bibfnamefont {R.~W.}\ \bibnamefont
  {Giannetta}}, \bibinfo {author} {\bibfnamefont {A.}~\bibnamefont
  {Carrington}}, \bibinfo {author} {\bibfnamefont {P.}~\bibnamefont
  {Fournier}}, \bibinfo {author} {\bibfnamefont {R.~L.}\ \bibnamefont
  {Greene}}, \bibinfo {author} {\bibfnamefont {P.}~\bibnamefont {Guptasarma}},
  \bibinfo {author} {\bibfnamefont {D.~G.}\ \bibnamefont {Hinks}},\ and\
  \bibinfo {author} {\bibfnamefont {A.~R.}\ \bibnamefont {Banks}},\ }\href {<Go
  to ISI>://WOS:000165824200041} {\bibfield  {journal} {\bibinfo  {journal}
  {Appl. Phys. Lett.}\ }\textbf {\bibinfo {volume} {77}},\ \bibinfo {pages}
  {4202} (\bibinfo {year} {2000}{\natexlab{b}})}\BibitemShut {NoStop}%
\bibitem [{\citenamefont {Prozorov}\ and\ \citenamefont
  {Giannetta}(2006)}]{Prozorov2006}%
  \BibitemOpen
  \bibfield  {author} {\bibinfo {author} {\bibfnamefont {R.}~\bibnamefont
  {Prozorov}}\ and\ \bibinfo {author} {\bibfnamefont {R.~W.}\ \bibnamefont
  {Giannetta}},\ }\href {https://doi.org/10.1088/0953-2048/19/8/r01} {\bibfield
   {journal} {\bibinfo  {journal} {Superconductor Science and Technology}\
  }\textbf {\bibinfo {volume} {19}},\ \bibinfo {pages} {R41} (\bibinfo {year}
  {2006})}\BibitemShut {NoStop}%
\bibitem [{\citenamefont {Prozorov}\ and\ \citenamefont
  {Kogan}(2011)}]{Prozorov2011}%
  \BibitemOpen
  \bibfield  {author} {\bibinfo {author} {\bibfnamefont {R.}~\bibnamefont
  {Prozorov}}\ and\ \bibinfo {author} {\bibfnamefont {V.~G.}\ \bibnamefont
  {Kogan}},\ }\href {http://stacks.iop.org/0034-4885/74/i=12/a=124505}
  {\bibfield  {journal} {\bibinfo  {journal} {Reports on Progress in Physics}\
  }\textbf {\bibinfo {volume} {74}},\ \bibinfo {pages} {124505} (\bibinfo
  {year} {2011})}\BibitemShut {NoStop}%
\bibitem [{\citenamefont {Carrington}(2011)}]{Carrington2011}%
  \BibitemOpen
  \bibfield  {author} {\bibinfo {author} {\bibfnamefont {A.}~\bibnamefont
  {Carrington}},\ }\href
  {https://doi.org/https://doi.org/10.1016/j.crhy.2011.03.001} {\bibfield
  {journal} {\bibinfo  {journal} {Comptes Rendus Physique}\ }\textbf {\bibinfo
  {volume} {12}},\ \bibinfo {pages} {502} (\bibinfo {year} {2011})},\ \bibinfo
  {note} {superconductivity of strongly correlated systems}\BibitemShut
  {NoStop}%
\bibitem [{\citenamefont {Prozorov}(2021)}]{Prozorov2021}%
  \BibitemOpen
  \bibfield  {author} {\bibinfo {author} {\bibfnamefont {R.}~\bibnamefont
  {Prozorov}},\ }\href {https://doi.org/10.1103/physrevapplied.16.024014}
  {\bibfield  {journal} {\bibinfo  {journal} {Physical Review Applied}\
  }\textbf {\bibinfo {volume} {16}},\ \bibinfo {pages} {024014} (\bibinfo
  {year} {2021})}\BibitemShut {NoStop}%
\bibitem [{\citenamefont {Prozorov}\ and\ \citenamefont
  {Kogan}(2018)}]{Prozorov2018}%
  \BibitemOpen
  \bibfield  {author} {\bibinfo {author} {\bibfnamefont {R.}~\bibnamefont
  {Prozorov}}\ and\ \bibinfo {author} {\bibfnamefont {V.~G.}\ \bibnamefont
  {Kogan}},\ }\href {https://doi.org/10.1103/PhysRevApplied.10.014030}
  {\bibfield  {journal} {\bibinfo  {journal} {Phys. Rev. Applied}\ }\textbf
  {\bibinfo {volume} {10}},\ \bibinfo {pages} {014030} (\bibinfo {year}
  {2018})}\BibitemShut {NoStop}%
\bibitem [{\citenamefont {Young}\ \emph {et~al.}(2005)\citenamefont {Young},
  \citenamefont {Moldovan}, \citenamefont {Adams},\ and\ \citenamefont
  {Prozorov}}]{Young2005}%
  \BibitemOpen
  \bibfield  {author} {\bibinfo {author} {\bibfnamefont {D.~P.}\ \bibnamefont
  {Young}}, \bibinfo {author} {\bibfnamefont {M.}~\bibnamefont {Moldovan}},
  \bibinfo {author} {\bibfnamefont {P.~W.}\ \bibnamefont {Adams}},\ and\
  \bibinfo {author} {\bibfnamefont {R.}~\bibnamefont {Prozorov}},\ }\href {<Go
  to ISI>://WOS:000229513300036
  http://iopscience.iop.org/0953-2048/18/5/034/pdf/0953-2048_18_5_034.pdf}
  {\bibfield  {journal} {\bibinfo  {journal} {Superc. Sci. Technol.}\ }\textbf
  {\bibinfo {volume} {18}},\ \bibinfo {pages} {776} (\bibinfo {year}
  {2005})}\BibitemShut {NoStop}%
\bibitem [{\citenamefont {Joshi}\ \emph {et~al.}(2022)\citenamefont {Joshi},
  \citenamefont {Ghimire}, \citenamefont {Tanatar}, \citenamefont {Datta},
  \citenamefont {Oh}, \citenamefont {Zhou}, \citenamefont {Kopas},
  \citenamefont {Marshall}, \citenamefont {Mutus}, \citenamefont {Slaughter},
  \citenamefont {Kramer}, \citenamefont {Sauls},\ and\ \citenamefont
  {Prozorov}}]{Joshi2022arXiv}%
  \BibitemOpen
  \bibfield  {author} {\bibinfo {author} {\bibfnamefont {K.~R.}\ \bibnamefont
  {Joshi}}, \bibinfo {author} {\bibfnamefont {S.}~\bibnamefont {Ghimire}},
  \bibinfo {author} {\bibfnamefont {M.~A.}\ \bibnamefont {Tanatar}}, \bibinfo
  {author} {\bibfnamefont {A.}~\bibnamefont {Datta}}, \bibinfo {author}
  {\bibfnamefont {J.-S.}\ \bibnamefont {Oh}}, \bibinfo {author} {\bibfnamefont
  {L.}~\bibnamefont {Zhou}}, \bibinfo {author} {\bibfnamefont {C.~J.}\
  \bibnamefont {Kopas}}, \bibinfo {author} {\bibfnamefont {J.}~\bibnamefont
  {Marshall}}, \bibinfo {author} {\bibfnamefont {J.~Y.}\ \bibnamefont {Mutus}},
  \bibinfo {author} {\bibfnamefont {J.}~\bibnamefont {Slaughter}}, \bibinfo
  {author} {\bibfnamefont {M.~J.}\ \bibnamefont {Kramer}}, \bibinfo {author}
  {\bibfnamefont {J.~A.}\ \bibnamefont {Sauls}},\ and\ \bibinfo {author}
  {\bibfnamefont {R.}~\bibnamefont {Prozorov}},\ }\href@noop {} {\bibfield
  {journal} {\bibinfo  {journal} {arXiv:2207.11616}\ } (\bibinfo {year}
  {2022})},\ \Eprint {https://arxiv.org/abs/2207.11616} {arXiv:2207.11616
  [cond-mat.supr-con]} \BibitemShut {NoStop}%
\bibitem [{\citenamefont {Tanatar}\ \emph {et~al.}(2022)\citenamefont
  {Tanatar}, \citenamefont {Torsello}, \citenamefont {Joshi}, \citenamefont
  {Ghimire}, \citenamefont {Zarea}, \citenamefont {Kopas}, \citenamefont
  {Ghigo}, \citenamefont {Sauls},\ and\ \citenamefont
  {Prozorov}}]{p-irr-bridge}%
  \BibitemOpen
  \bibfield  {author} {\bibinfo {author} {\bibfnamefont {M.~A.}\ \bibnamefont
  {Tanatar}}, \bibinfo {author} {\bibfnamefont {D.}~\bibnamefont {Torsello}},
  \bibinfo {author} {\bibfnamefont {K.~R.}\ \bibnamefont {Joshi}}, \bibinfo
  {author} {\bibfnamefont {S.}~\bibnamefont {Ghimire}}, \bibinfo {author}
  {\bibfnamefont {M.}~\bibnamefont {Zarea}}, \bibinfo {author} {\bibfnamefont
  {C.~J.}\ \bibnamefont {Kopas}}, \bibinfo {author} {\bibfnamefont
  {G.}~\bibnamefont {Ghigo}}, \bibinfo {author} {\bibfnamefont {J.~A.}\
  \bibnamefont {Sauls}},\ and\ \bibinfo {author} {\bibfnamefont
  {R.}~\bibnamefont {Prozorov}},\ }\href@noop {} {\bibfield  {journal}
  {\bibinfo  {journal} {in preparation}\ } (\bibinfo {year}
  {2022})}\BibitemShut {NoStop}%
\bibitem [{\citenamefont {Prozorov}\ \emph {et~al.}(2001)\citenamefont
  {Prozorov}, \citenamefont {Giannetta}, \citenamefont {Bud'ko},\ and\
  \citenamefont {Canfield}}]{Prozorov_Proximity}%
  \BibitemOpen
  \bibfield  {author} {\bibinfo {author} {\bibfnamefont {R.}~\bibnamefont
  {Prozorov}}, \bibinfo {author} {\bibfnamefont {R.~W.}\ \bibnamefont
  {Giannetta}}, \bibinfo {author} {\bibfnamefont {S.~L.}\ \bibnamefont
  {Bud'ko}},\ and\ \bibinfo {author} {\bibfnamefont {P.~C.}\ \bibnamefont
  {Canfield}},\ }\href {https://doi.org/10.1103/PhysRevB.64.180501} {\bibfield
  {journal} {\bibinfo  {journal} {Phys. Rev. B}\ }\textbf {\bibinfo {volume}
  {64}},\ \bibinfo {pages} {180501} (\bibinfo {year} {2001})}\BibitemShut
  {NoStop}%
\bibitem [{\citenamefont {Pambianchi}\ \emph {et~al.}(1995)\citenamefont
  {Pambianchi}, \citenamefont {Mao},\ and\ \citenamefont
  {Anlage}}]{Analage_1995}%
  \BibitemOpen
  \bibfield  {author} {\bibinfo {author} {\bibfnamefont {M.~S.}\ \bibnamefont
  {Pambianchi}}, \bibinfo {author} {\bibfnamefont {S.~N.}\ \bibnamefont
  {Mao}},\ and\ \bibinfo {author} {\bibfnamefont {S.~M.}\ \bibnamefont
  {Anlage}},\ }\href {https://doi.org/10.1103/PhysRevB.52.4477} {\bibfield
  {journal} {\bibinfo  {journal} {Phys. Rev. B}\ }\textbf {\bibinfo {volume}
  {52}},\ \bibinfo {pages} {4477} (\bibinfo {year} {1995})}\BibitemShut
  {NoStop}%
\bibitem [{\citenamefont {Roth}\ \emph {et~al.}(1990)\citenamefont {Roth},
  \citenamefont {Kurakin}, \citenamefont {Piel}, \citenamefont {Heinrichs},\
  and\ \citenamefont {Pouryamout}}]{roth1990suppression}%
  \BibitemOpen
  \bibfield  {author} {\bibinfo {author} {\bibfnamefont {R.}~\bibnamefont
  {Roth}}, \bibinfo {author} {\bibfnamefont {V.}~\bibnamefont {Kurakin}},
  \bibinfo {author} {\bibfnamefont {H.}~\bibnamefont {Piel}}, \bibinfo {author}
  {\bibfnamefont {H.}~\bibnamefont {Heinrichs}},\ and\ \bibinfo {author}
  {\bibfnamefont {J.}~\bibnamefont {Pouryamout}},\ }\href@noop {} {\  (\bibinfo
  {year} {1990})}\BibitemShut {NoStop}%
\bibitem [{\citenamefont {Berti}\ \emph {et~al.}(2023)\citenamefont {Berti},
  \citenamefont {Torres-Castanedo}, \citenamefont {Goronzy}, \citenamefont
  {Bedzyk}, \citenamefont {Hersam}, \citenamefont {Kopas}, \citenamefont
  {Marshall},\ and\ \citenamefont {Lavarone}}]{berti2023scanning}%
  \BibitemOpen
  \bibfield  {author} {\bibinfo {author} {\bibfnamefont {G.}~\bibnamefont
  {Berti}}, \bibinfo {author} {\bibfnamefont {C.}~\bibnamefont
  {Torres-Castanedo}}, \bibinfo {author} {\bibfnamefont {D.}~\bibnamefont
  {Goronzy}}, \bibinfo {author} {\bibfnamefont {M.}~\bibnamefont {Bedzyk}},
  \bibinfo {author} {\bibfnamefont {M.}~\bibnamefont {Hersam}}, \bibinfo
  {author} {\bibfnamefont {C.}~\bibnamefont {Kopas}}, \bibinfo {author}
  {\bibfnamefont {J.}~\bibnamefont {Marshall}},\ and\ \bibinfo {author}
  {\bibfnamefont {M.}~\bibnamefont {Lavarone}},\ }\href@noop {} {\bibfield
  {journal} {\bibinfo  {journal} {Applied Physics Letters}\ }\textbf {\bibinfo
  {volume} {122}} (\bibinfo {year} {2023})}\BibitemShut {NoStop}%
\bibitem [{\citenamefont {Chandrasekhar}\ and\ \citenamefont
  {Einzel}(1993)}]{Chandrasekhar1993}%
  \BibitemOpen
  \bibfield  {author} {\bibinfo {author} {\bibfnamefont {B.~S.}\ \bibnamefont
  {Chandrasekhar}}\ and\ \bibinfo {author} {\bibfnamefont {D.}~\bibnamefont
  {Einzel}},\ }\href {https://doi.org/https://doi.org/10.1002/andp.19935050604}
  {\bibfield  {journal} {\bibinfo  {journal} {Annalen der Physik}\ }\textbf
  {\bibinfo {volume} {505}},\ \bibinfo {pages} {535} (\bibinfo {year}
  {1993})}\BibitemShut {NoStop}%
\bibitem [{\citenamefont {Einzel}\ \emph {et~al.}(1986)\citenamefont {Einzel},
  \citenamefont {Hirschfeld}, \citenamefont {Gross}, \citenamefont
  {Chandrasekhar}, \citenamefont {Andres}, \citenamefont {Ott}, \citenamefont
  {Beuers}, \citenamefont {Fisk},\ and\ \citenamefont {Smith}}]{Einzel1986}%
  \BibitemOpen
  \bibfield  {author} {\bibinfo {author} {\bibfnamefont {D.}~\bibnamefont
  {Einzel}}, \bibinfo {author} {\bibfnamefont {P.~J.}\ \bibnamefont
  {Hirschfeld}}, \bibinfo {author} {\bibfnamefont {F.}~\bibnamefont {Gross}},
  \bibinfo {author} {\bibfnamefont {B.~S.}\ \bibnamefont {Chandrasekhar}},
  \bibinfo {author} {\bibfnamefont {K.}~\bibnamefont {Andres}}, \bibinfo
  {author} {\bibfnamefont {H.~R.}\ \bibnamefont {Ott}}, \bibinfo {author}
  {\bibfnamefont {J.}~\bibnamefont {Beuers}}, \bibinfo {author} {\bibfnamefont
  {Z.}~\bibnamefont {Fisk}},\ and\ \bibinfo {author} {\bibfnamefont {J.~L.}\
  \bibnamefont {Smith}},\ }\href {https://doi.org/10.1103/PhysRevLett.56.2513}
  {\bibfield  {journal} {\bibinfo  {journal} {Phys. Rev. Lett.}\ }\textbf
  {\bibinfo {volume} {56}},\ \bibinfo {pages} {2513} (\bibinfo {year}
  {1986})}\BibitemShut {NoStop}%
\bibitem [{\citenamefont {Einzel}(2003)}]{Einzel2003}%
  \BibitemOpen
  \bibfield  {author} {\bibinfo {author} {\bibfnamefont {D.}~\bibnamefont
  {Einzel}},\ }\href {https://doi.org/10.1023/A:1022872911344} {\bibfield
  {journal} {\bibinfo  {journal} {Journal of Low Temperature Physics}\ }\textbf
  {\bibinfo {volume} {131}},\ \bibinfo {pages} {1} (\bibinfo {year}
  {2003})}\BibitemShut {NoStop}%
\bibitem [{\citenamefont {Gurevich}(2023)}]{Gurevich2023}%
  \BibitemOpen
  \bibfield  {author} {\bibinfo {author} {\bibfnamefont {A.~V.}\ \bibnamefont
  {Gurevich}},\ }\bibfield  {journal} {\bibinfo  {journal} {Supercond. Sci.
  Technol.}\ }\href {https://doi.org/10.1088/1361-6668/acc214}
  {10.1088/1361-6668/acc214} (\bibinfo {year} {2023})\BibitemShut {NoStop}%
\bibitem [{\citenamefont {Gurevich}\ and\ \citenamefont
  {Kubo}(2017)}]{Gurevich2017}%
  \BibitemOpen
  \bibfield  {author} {\bibinfo {author} {\bibfnamefont {A.}~\bibnamefont
  {Gurevich}}\ and\ \bibinfo {author} {\bibfnamefont {T.}~\bibnamefont
  {Kubo}},\ }\href {https://doi.org/10.1103/PhysRevB.96.184515} {\bibfield
  {journal} {\bibinfo  {journal} {Phys. Rev. B}\ }\textbf {\bibinfo {volume}
  {96}},\ \bibinfo {pages} {184515} (\bibinfo {year} {2017})}\BibitemShut
  {NoStop}%
\bibitem [{\citenamefont {Kubo}(2020{\natexlab{a}})}]{Kubo2020}%
  \BibitemOpen
  \bibfield  {author} {\bibinfo {author} {\bibfnamefont {T.}~\bibnamefont
  {Kubo}},\ }\href {https://doi.org/10.1103/physrevresearch.2.033203}
  {\bibfield  {journal} {\bibinfo  {journal} {Phys. Rev. Res.}\ }\textbf
  {\bibinfo {volume} {2}},\ \bibinfo {pages} {33203} (\bibinfo {year}
  {2020}{\natexlab{a}})}\BibitemShut {NoStop}%
\bibitem [{\citenamefont {Kubo}(2020{\natexlab{b}})}]{Kubo2020a}%
  \BibitemOpen
  \bibfield  {author} {\bibinfo {author} {\bibfnamefont {T.}~\bibnamefont
  {Kubo}},\ }\bibfield  {journal} {\bibinfo  {journal} {Phys. Rev. Res.}\
  }\textbf {\bibinfo {volume} {2}},\ \href
  {https://doi.org/10.1103/PhysRevResearch.2.013302}
  {10.1103/PhysRevResearch.2.013302} (\bibinfo {year}
  {2020}{\natexlab{b}})\BibitemShut {NoStop}%
\bibitem [{\citenamefont {Herman}\ and\ \citenamefont
  {Hlubina}(2016)}]{Herman2016}%
  \BibitemOpen
  \bibfield  {author} {\bibinfo {author} {\bibfnamefont {F.}~\bibnamefont
  {Herman}}\ and\ \bibinfo {author} {\bibfnamefont {R.}~\bibnamefont
  {Hlubina}},\ }\href {https://doi.org/10.1103/PhysRevB.94.144508} {\bibfield
  {journal} {\bibinfo  {journal} {Phys. Rev. B}\ }\textbf {\bibinfo {volume}
  {94}},\ \bibinfo {pages} {1} (\bibinfo {year} {2016})}\BibitemShut {NoStop}%
\bibitem [{\citenamefont {Herman}\ and\ \citenamefont
  {Hlubina}(2017)}]{Herman2017}%
  \BibitemOpen
  \bibfield  {author} {\bibinfo {author} {\bibfnamefont {F.}~\bibnamefont
  {Herman}}\ and\ \bibinfo {author} {\bibfnamefont {R.}~\bibnamefont
  {Hlubina}},\ }\href {https://doi.org/10.1103/PhysRevB.96.014509} {\bibfield
  {journal} {\bibinfo  {journal} {Phys. Rev. B}\ }\textbf {\bibinfo {volume}
  {96}},\ \bibinfo {pages} {1} (\bibinfo {year} {2017})}\BibitemShut {NoStop}%
\bibitem [{\citenamefont {Herman}\ and\ \citenamefont
  {Hlubina}(2018)}]{Herman2018}%
  \BibitemOpen
  \bibfield  {author} {\bibinfo {author} {\bibfnamefont {F.}~\bibnamefont
  {Herman}}\ and\ \bibinfo {author} {\bibfnamefont {R.}~\bibnamefont
  {Hlubina}},\ }\href {https://doi.org/10.1103/PhysRevB.97.014517} {\bibfield
  {journal} {\bibinfo  {journal} {Phys. Rev. B}\ }\textbf {\bibinfo {volume}
  {97}},\ \bibinfo {pages} {1} (\bibinfo {year} {2018})}\BibitemShut {NoStop}%
\bibitem [{\citenamefont {Lechner}\ \emph {et~al.}(2020)\citenamefont
  {Lechner}, \citenamefont {Oli}, \citenamefont {Makita}, \citenamefont
  {Ciovati}, \citenamefont {Gurevich},\ and\ \citenamefont
  {Iavarone}}]{Lechner_PRA_2020}%
  \BibitemOpen
  \bibfield  {author} {\bibinfo {author} {\bibfnamefont {E.~M.}\ \bibnamefont
  {Lechner}}, \bibinfo {author} {\bibfnamefont {B.~D.}\ \bibnamefont {Oli}},
  \bibinfo {author} {\bibfnamefont {J.}~\bibnamefont {Makita}}, \bibinfo
  {author} {\bibfnamefont {G.}~\bibnamefont {Ciovati}}, \bibinfo {author}
  {\bibfnamefont {A.}~\bibnamefont {Gurevich}},\ and\ \bibinfo {author}
  {\bibfnamefont {M.}~\bibnamefont {Iavarone}},\ }\href
  {https://doi.org/10.1103/PhysRevApplied.13.044044} {\bibfield  {journal}
  {\bibinfo  {journal} {Phys. Rev. Appl.}\ }\textbf {\bibinfo {volume} {13}},\
  \bibinfo {pages} {044044} (\bibinfo {year} {2020})}\BibitemShut {NoStop}%
\bibitem [{\citenamefont {Abrikosov}\ and\ \citenamefont
  {Gor'kov}(1960)}]{AbrikosovGorkov1960ZETF}%
  \BibitemOpen
  \bibfield  {author} {\bibinfo {author} {\bibfnamefont {A.~A.}\ \bibnamefont
  {Abrikosov}}\ and\ \bibinfo {author} {\bibfnamefont {L.~P.}\ \bibnamefont
  {Gor'kov}},\ }\href@noop {} {\bibfield  {journal} {\bibinfo  {journal} {Zh.
  Eksp. Teor. Fiz. (Sov. Phys. JETP 12, 1243 (1961))}\ }\textbf {\bibinfo
  {volume} {39}},\ \bibinfo {pages} {1781} (\bibinfo {year}
  {1960})}\BibitemShut {NoStop}%
\bibitem [{\citenamefont {Gurevich}(2003)}]{Gurevich2003}%
  \BibitemOpen
  \bibfield  {author} {\bibinfo {author} {\bibfnamefont {A.}~\bibnamefont
  {Gurevich}},\ }\href {https://doi.org/10.1103/physrevb.67.184515} {\bibfield
  {journal} {\bibinfo  {journal} {Physical Review B}\ }\textbf {\bibinfo
  {volume} {67}},\ \bibinfo {pages} {184515} (\bibinfo {year}
  {2003})}\BibitemShut {NoStop}%
\end{thebibliography}

%

\end{document}